\documentclass[prd,nofootinbib,twocolumn,superscriptaddress]{revtex4}
\usepackage{hyperref}
\usepackage{placeins}
\usepackage{graphicx}
\usepackage[usenames, dvipsnames]{color}
\usepackage{soul}
\usepackage{comment}
\usepackage[normalem]{ulem}
\usepackage{amsmath, amssymb, bbold, mathrsfs, slashed}
\usepackage{multirow, bm, IEEEtrantools}
\usepackage{enumitem}
\usepackage{tabularx}

\usepackage{orcidlink}
\usepackage{acro}

\DeclareAcronym{tov}{
  short=TOV,
  long=Tolman--Oppenheimer--Volkoff,
}

\DeclareAcronym{sm}{
  short=SM,
  long=standard model,
}

\DeclareAcronym{ns}{
  short=NS,
  long=neutron star,
}

\DeclareAcronym{hs}{
  short=HS,
  long=hybrid star,
}

\DeclareAcronym{qcd}{
  short=QCD,
  long=quantum chromodynamics,
}

\DeclareAcronym{pqcd}{
  short=pQCD,
  long=perturbative quantum chromodynamics,
}

\DeclareAcronym{lqcd}{
  short=lQCD,
  long=lattice quantum chromodynamics,
}

\DeclareAcronym{eos}{
  short=EoS,
  long=equation of state,
}

\DeclareAcronym{nsm}{
  short=NSM,
  long=neutron star matter,
}

\DeclareAcronym{nm}{
  short=NM,
  long=nuclear matter,
}

\DeclareAcronym{ddb}{
  short=DDB,
  long=density-dependent couplings with Bayesian analysis,
}

\DeclareAcronym{rmf}{
  short=RMF,
  long=relativistic mean field,
}

\DeclareAcronym{nro}{
  short=NRO,
  long=non-radial oscillation,
}

\DeclareAcronym{ai}{
  short=AI,
  long=artificial intelligence,
}

\DeclareAcronym{gw}{
  short=GW,
  long=gravitational wave,
}

\DeclareAcronym{gr}{
  short=GR,
  long=general relativity,
}

\DeclareAcronym{nicer}{
  short=NICER,
  long=Neutron Star Interior Composition ExploreR,
}

\DeclareAcronym{hp}{
  short=HP,
  long=hadronic phase,
}

\DeclareAcronym{mp}{
  short=MP,
  long=mixed phase,
}

\DeclareAcronym{qp}{
  short=QP,
  long=quark phase,
}

\DeclareAcronym{njl}{
  short=NJL,
  long=Nambu--Jona-Lasinio,
}

\DeclareAcronym{ml}{
  short=ML,
  long=machine learning,
}

\DeclareAcronym{nl}{
  short=NL,
  long=non-linear,
}

\DeclareAcronym{pca}{
  short=PCA,
  long=principal component analysis,
}

\DeclareAcronym{qnm}{
  short=QNM,
  long=quasi-normal mode,
}
\DeclareAcronym{dm}{
  short=DM,
  long=dark matter,
}
\DeclareAcronym{adm}{
  short=ADM,
  long= admixed dark matter,
}

\begin{document}

\title{Neutron stars with an agnostic Dark sector: Core and Halo configurations from a two-fluid approach}

\author{Asit Karan\orcidlink{0009-0005-3070-2729}}
\email{asit22@iiserb.ac.in}
\affiliation{Department of Physics, Indian Institute of Science Education and Research, Bhopal 462066, India}

\author{Tuhin Malik\orcidlink{0000-0003-2633-5821}}
\email{tm@uc.pt}
\affiliation{CFisUC, Department of Physics, University of Coimbra, 3004-516 Coimbra, Portugal}

\author{Asim Kumar Saha\orcidlink{0009-0000-8375-4833}}
\email{asim21@iiserb.ac.in}
\affiliation{Department of Physics, Indian Institute of Science Education and Research, Bhopal 462066, India}

\author{Constança Providência\orcidlink{0000-0001-6464-8023}}
\email{cp@uc.pt}
\affiliation{CFisUC, Department of Physics, University of Coimbra, 3004-516 Coimbra, Portugal}

\author{Ritam Mallick\orcidlink{0000-0003-2943-6388}}
\email{mallick@iiserb.ac.in}
\affiliation{Department of Physics, Indian Institute of Science Education and Research, Bhopal 462066, India}

% https://orcid.org/0009-0000-8375-4833 AKS
% https://orcid.org/0000-0003-2633-5821 TM
% https://orcid.org/0000-0001-6464-8023 CP

\begin{abstract}

 The study of dark matter admixed neutron stars has the potential to advance our understanding of dark matter particle candidates. However, the large parameter space of dark matter particle masses restricts a systematic, model-independent study. In this analysis, we employ agnostic hadronic and dark-matter equations of state to construct dark-matter-admixed neutron stars within a two-fluid formalism. Dark matter is characterized solely by its low-density equation of state and mass and is modeled as a Fermi gas, while hadronic matter is anchored at low and high densities by chiral effective field theory and perturbative quantum chromodynamics calculations. A speed-of-sound parametrization describes the intermediate- and high-density regimes of both hadronic and dark matter, enabling an agnostic treatment of the dark matter equation of state that is constrained only by thermodynamic consistency and does not favor softer or stiffer behavior. Within this agnostic framework, we find that dark matter does not generically compactify the star: light dark matter forms extended halos that raise the tidal deformability, while heavy dark matter forms compact cores that lower it. Consequently, the dominant observational constraint shifts from gravitational-wave tidal deformability for light, halo-dominated models to NICER mass-radius data for heavy, core-dominated models. Using current data at $1\sigma$, we constrain the dark matter fraction to $f_{\mathrm{DM}} \lesssim 0.11$ for light dark matter. By treating the medium- and high-density dark matter equation of state agnostically within a fermionic framework, our approach yields conservative and broadly applicable bounds on the dark matter content of neutron stars. Neutron stars with similar masses but very different tidal deformabilities could be a smoking-gun signature of dark matter in neutron stars.

\end{abstract}

\maketitle
%\flushbottom

\section{Introduction}
Neutron stars (NSs) are among the densest objects in the Universe, with central densities reaching several times the nuclear saturation density $n_0 \simeq 0.16~\mathrm{fm}^{-3}$ and gravitational fields surpassed only by black holes. These extreme conditions render them unrivaled natural laboratories for probing fundamental physics: the behavior of strongly interacting matter at supra-nuclear densities, the validity of general relativity in the strong-field regime, and a wealth of phenomena beyond the Standard Model that cannot be accessed in terrestrial experiments~\cite{LattimerPrakash2004,LattimerPrakash2007,OzelFreire2016,Baym2018,Burgio2021}. The global structure of an NS—in particular its mass $M$, radius $R$, and tidal deformability $\Lambda$—is fixed by the equation of state (EoS) of cold, charge-neutral, $\beta$-equilibrated matter, so that each precise astrophysical measurement maps directly onto a constraint on the microphysics of dense matter.

Despite decades of effort, the dense-matter EoS remains the central unknown of NS physics. No single theoretical framework is reliable across the full density range realized inside a NS: chiral effective field theory (CET) provides a systematic, controlled description of nuclear matter only near and slightly above $n_0$~\cite{HebelerSchwenk2010,Tews2013,Drischler2019,Drischler2021b}, while perturbative quantum chromodynamics (pQCD) becomes trustworthy only at asymptotically high densities, $n \gtrsim 40\,n_0$~\cite{Kurkela2010,Fraga:2013qra,Gorda2018,Komoltsev2022,Gorda2023}. In between lies precisely the regime that governs NS interiors, where the relevant degrees of freedom---nucleons, hyperons, or deconfined quarks---are themselves uncertain~\cite{LattimerPrakash2016,Oertel2017,Baym2018}. EoSs built on a fixed microscopic Lagrangian therefore risk biasing the inferred stellar properties toward assumptions that cannot be independently validated.

This impasse has motivated a now-standard \emph{agnostic} strategy: anchor the EoS to controlled calculations at the boundaries of the uncertain region---CET at low density and pQCD at high density---and interpolate between them with a flexible parametrization required only to be thermodynamically stable and causal. Several such schemes are in common use, including piecewise polytropes~\cite{Read2009,Hebeler_2013,Kurkela_2014}, speed-of-sound interpolation~\cite{Tews2018,Greif2019,Raithel2016,Annala:2019puf,Altiparmak:2022,Tan2022,Brandes2023, Nath:2023gmu, Verma:2025dez}, nonparametric Gaussian processes~\cite{Landry2019,Essick2020,Legred2021}, and Taylor-expansion (nuclear meta-modeling) about saturation~\cite{Margueron2018a,Margueron2018b}. Large agnostic EoS ensembles constructed in this way have become the backbone of modern dense-matter inference~\cite{Annala2018,Most2018,Annala2022,Annala2023,sagnik_2024,anshuman_2025,Saha:2024swd,thakur_2024,asim_2026}. In the present work, we adopt the multi-segment speed-of-sound interpolation of Ref.~\cite{Annala:2019puf} for the nuclear sector, matched to a Baym--Pethick--Sutherland crust~\cite{BPS} and a CET band up to $\sim\!1.1\,n_0$, and anchored to the pQCD limit at high density.

The agnostic program is powerful precisely because the dense-matter EoS is now confronted by an increasingly sharp multi-messenger observational dataset. Radio timing of binary millisecond pulsars has established the existence of two-solar-mass NSs---PSR~J1614$-$2230~\cite{Demorest2010}, PSR~J0348$+$0432 at $2.01\pm0.04\,M_\odot$~\cite{Antoniadis2013}, and the high-mass PSR~J0740$+$6620~\cite{Cromartie2020,Fonseca2021}---with the black-widow pulsar PSR~J0952$-$0607 reaching $2.35\pm0.11\,M_\odot$~\cite{Romani:2025ytn}, imposing a stringent lower bound on the maximum mass that any viable EoS must support. NASA's Neutron Star Interior Composition Explorer (NICER) has delivered simultaneous mass-radius measurements through pulse-profile modeling of PSR~J0030$+$0451~\cite{Riley2019,Miller2019}, PSR~J0740$+$6620~\cite{Riley2021,Miller2021,Salmi2024,Dittmann2024}, the nearby PSR~J0437$-$4715~\cite{Choudhury2024}, and most recently PSR~J0614$-$3329~\cite{Mauviard2025}. Finally, the gravitational-wave detection of the binary NS merger GW170817~\cite{GW170817disc} opened tidal deformability as a direct EoS probe, constraining the dimensionless tidal deformability of a $1.4\,M_\odot$ star to $70 \leq \Lambda_{1.4} \leq 580$~\cite{GW170817,GW170817prop}.

Although NS in itself serves as a unique laboratory to test matter properties, going one step further, it can even serve as a laboratory for the search for dark matter (DM). The DM constitutes roughly a quarter of the cosmic energy budget~\cite{Planck2020} yet has so far revealed itself only through its gravitational influence. NSs offer a unique window onto the dark sector: over their lifetimes, they can gravitationally capture and accrete DM particles from their surrounding environment, a possibility first explored for compact stars in Refs.~\cite{PressSpergel1985,GoldmanNussinov1989} and developed into a rich program of constraints on the DM mass and interaction cross-section~\cite{Kouvaris2008,BertoneFairbairn2008,deLavallazFairbairn2010,KouvarisTinyakov2011,Bell2020}. Depending on its mass and interaction strength, captured DM may accumulate within the stellar interior and measurably alter the star's macroscopic properties, so that realistic NSs may in fact be dark matter-admixed neutron stars (DMANSs).

When the dark and visible sectors interact only gravitationally, a DMANS is naturally described as a two-fluid system: two coupled Tolman--Oppenheimer--Volkoff (TOV) equations that share a common metric but carry separate pressures and energy densities~\cite{SandinCiarcelluti2009,CiarcellutiSandin2011,LeungChuLin2011,Goldman2013}. A central result of this picture is that the spatial distribution of the DM controls its observational imprint: DM confined within the nucleonic radius forms a dense \emph{core} that increases stellar compactness and reduces tidal deformability, whereas DM extending beyond the nucleonic surface forms a dilute \emph{halo} that enlarges the effective radius and enhances $\Lambda$.

The structural and observational consequences of admixed DM have been studied extensively, but almost always under the assumption of a \emph{specific} microscopic DM model. Bosonic DM, supported by self-interaction or quantum pressure, has been treated as both core- and halo-forming~\cite{NelsonReddyZhou2019,Karkevandi2022,Giangrandi2023}; fermionic DM, supported by degeneracy pressure, has been studied through free or self-interacting dark Fermi gases~\cite{NarainSchaffnerBielich2006,TolosSchaffnerBielich2015,Mukhopadhyay2016,IvanytskyiSagunLopes2020,Scordino:2024ehe, Barbat:2024yvi, Mariani:2023wtv, Vikiaris:2023vau, Rutherford:2024uix, Guha:2024pnn, karan:2025kel, Kumar:2025oyx, Kumar:2026hoq}; and a large body of work couples DM to nucleons directly, for example, through a Higgs portal, yielding single-fluid EoSs~\cite{PanotopoulosLopes2017,DasMalikNayak2019,DasMalikNayak2022,SenGuha2021,Lenzi2023,Lourenco2022,Routaray2023}. The tidal-deformability and mass-radius signatures of such configurations, and the resulting observational bounds, have likewise been worked out for particular candidates~\cite{LeungChuLin2022,CollierCroonLeane2022,Diedrichs2023,Rutherford2023,Mariani:2023wtv,ShirkeGhosh2023}. In every case, however, the dark EoS is fixed by a chosen particle nature, mass, and set of couplings, so that the inferred stellar signatures are inextricably tied to a narrow particle-physics hypothesis.

This is precisely the situation that, in the nuclear sector, motivated the shift toward agnostic EoS reconstruction in the first place. In this work, we apply the same agnostic philosophy to the dark sector. Rather than committing to a bosonic, fermionic, or portal-coupled candidate, we construct the DM EoS using the identical speed-of-sound interpolation scheme employed for nuclear matter: the low-density regime ($n < 0.1\,n_0$) is described by a free Fermi gas parametrized solely by a bare DM particle mass $m_D$, above which the EoS is generated by an agnostic, randomized speed-of-sound profile. The crucial distinction from the nuclear case is that DM possesses no analogous high-density (pQCD) anchor, so its agnostic interpolation is bounded only from below. To the best of our knowledge, this is the first comprehensive study in which \emph{both} the nuclear and the dark EoSs are reconstructed agnostically within a single, consistent framework, enabling the generic imprint of a dark component to be disentangled from model-specific assumptions.

Equipped with this framework, we solve the two-fluid TOV equations together with the tidal-deformability equations to generate $\sim\!10^5$ mass-radius sequences for each of several DM models spanning particle masses $m_D = 0.2$--$1.1~\mathrm{GeV}$ and DM mass fractions $f_{\mathrm{DM}} = 0.01$--$0.15$, and confront them with the full multi-messenger dataset: the NICER mass-radius measurements, the GW170817 bound $70 \leq \Lambda_{1.4} \leq 580$, and the maximum-mass requirement $M_{\mathrm{max}} \geq 2.01\,M_\odot$. We find that admixed DM softens the effective EoS and compactifies the star at low and intermediate masses, while for certain dark EoSs it can instead stiffen the configuration and raise the maximum mass; light DM preferentially forms extended halos that are most stringently constrained by the tidal deformability, whereas heavy DM forms compact cores most stringently constrained by the NICER mass-radius data. The remainder of this paper is organized as follows. In Sec .~II, we describe the theoretical formalism---the agnostic construction of the nuclear and dark EoSs, the two-fluid TOV equations, and the tidal-deformability formalism. In Sec .~III, we present and discuss our results before summarizing our main findings and their implications for the dark matter content of neutron stars.

\section{Theoretical Formalism}
In this section, we outline the theoretical framework employed in our analysis. We first construct the equation of state for nuclear matter (NM) and extend it to include a dark matter component. The equilibrium structure of dark matter-admixed neutron stars is then described within the two-fluid TOV formalism, in which the nuclear matter and dark matter components are treated as distinct fluids that interact only through gravity. In addition, we incorporate the equations governing tidal deformability, which enable direct comparison of our results with astrophysical observations.

\subsection{EoS Construction}\label{eos_construction}
\begin{figure*}[ht]
    \includegraphics[width=0.47\linewidth]{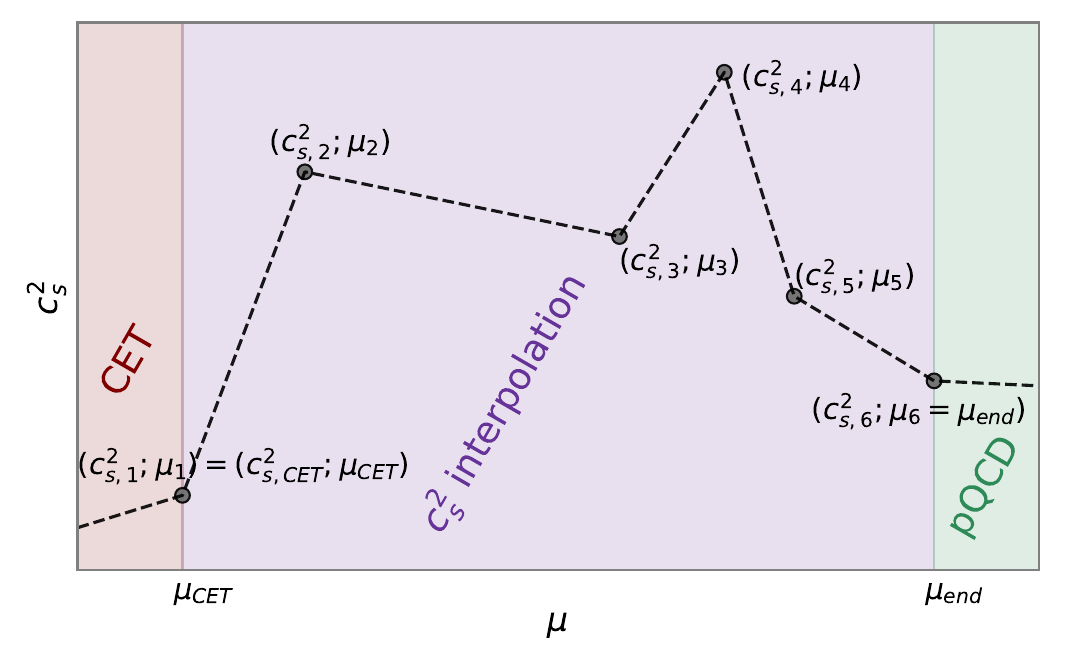}
    \includegraphics[width=0.47\linewidth]{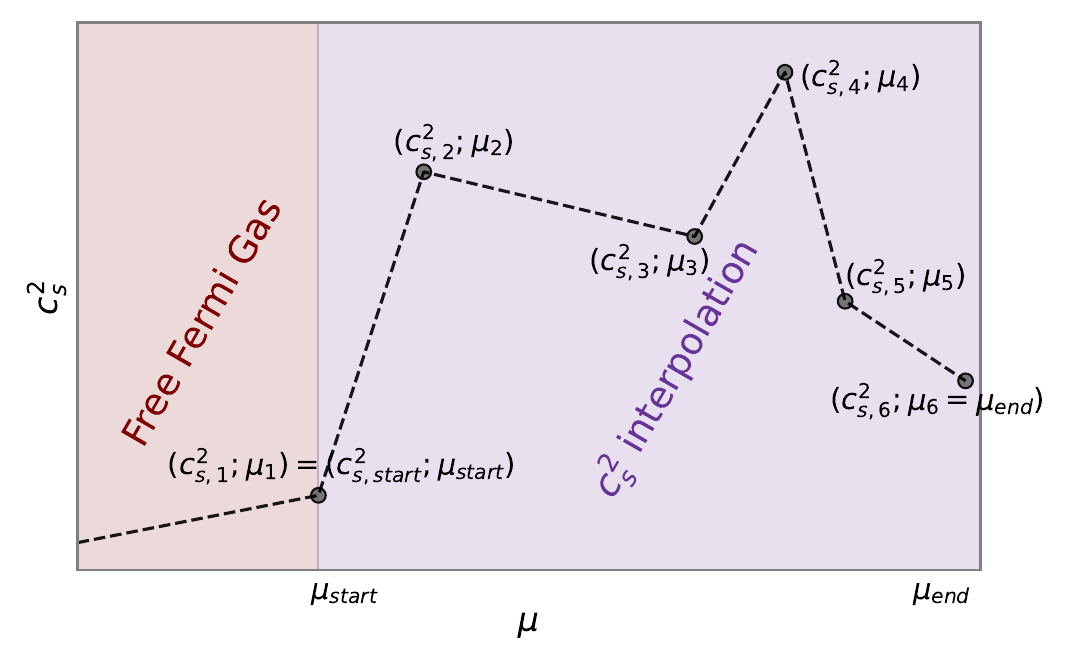}
    \caption{Schematic illustration of the interpolation scheme, highlighting the anchoring of the nuclear matter (NM) and dark matter (DM) EoS parameters at opposite ends of the interpolation interval. The left and right panels correspond to the NM and DM interpolation schemes, respectively.}
    \label{sosischematic}
\end{figure*}

We adopt the speed-of-sound interpolation method for our agnostic modeling of the EoS of both nuclear matter and dark matter. \textcolor{black}{Several alternative agnostic approaches have been developed in the literature, including piecewise-polytropic parameterizations, spectral representations, and Gaussian-process regression, which have been shown to yield comparable results \cite{Annala:2019puf}.} Fig.~\ref{sosischematic} illustrates the difference in interpolation schemes adopted for nuclear matter (left panel) and dark matter (right panel). The key distinction between the two lies in the availability of anchor points that serve as boundary conditions for integrating Eqs.~\ref{1} and \ref{2}. For the NM EoS, the interpolation is constrained at both low and high densities. At low densities, the anchor point is provided by the CET band at $1.1\,n_0$, while at high densities, an additional anchor is imposed from the onset of the pQCD regime. Consequently, the NM EoS interpolation is bounded from both ends. In contrast, for the DM EoS, only a low-density anchor point is available, chosen at $0.1\,n_0$, below which the matter is modeled as a free Fermi gas. Unlike the NM case, there exists no analogous asymptotic anchor for dark matter at high densities.

\textcolor{black}{Since the NM EoS and DM EoS share the same agnostic modeling, we shall briefly discuss the method here. The speed of sound interpolation (SOSI) method parametrizes the speed of sound ($c_s^2$) as a function of chemical potential ($\mu$), which allows us to write the number density as} 
\begin{equation}
n(\mu) = n_{\scriptscriptstyle start}\exp\left[\int_{\mu_{start}}^{\mu} \frac{d\mu'}{\mu' c_s^2(\mu')}\right]
\label{1}
\end{equation}
\textcolor{black}{where $n_{start}$ is fixed by matching the low-density anchor (CET for NM, and free Fermi gas for DM) to the initial values of the interpolated EoS, and $\mu_{start}$ is the chemical potential corresponding to $n_{start}$. Pressure can then readily be calculated as}
\begin{equation}
P(\mu) = P_{\scriptscriptstyle start} \! + n_{\scriptscriptstyle start}\int_{\mu_{start}}^{\mu}d\mu' \exp\left[\int_{\mu_{start}}^{\mu'}\frac{d\mu''}{\mu''c_s^2(\mu'')}\right]
\label{2}
\end{equation}   
\textcolor{black}{where $P_{start}$ is the pressure at $n_{start}$. The interpolation employs five randomized segments, $(c_{s,i}^2,\mu_i)$, with each $\mu_i$ sampled randomly from the interval $[\mu_{\scriptscriptstyle start},2.6~\mathrm{GeV}]$. For the NM EoS, the upper limit of $\mu$ is chosen such that the uncertainty in the pQCD regime is comparable to that at the CET \cite{Kurkela_2014}. The same upper limit is adopted for the DM EoS to ensure that its high-density regime spans the same range as that of the NM EoS. A piecewise linear function is employed to connect the points $\left\{\mu_{i},c_{s,i}^{2}\right\}$ as:}
\begin{equation}
c_s^2(\mu) = \frac{(\mu_{i+1} - \mu)c_{s,i}^2 + (\mu - \mu_{i})c_{s,i+1}^2}{\mu_{i+1} - \mu_i}
\label{3}
\end{equation}
\textcolor{black}{where $c_{s,i}^2 \in [0,1]$. This function allows us to carry out the integrals in \eqref{1} and \eqref{2}.} 

\subsubsection{Nuclear matter EoS}
We construct our high-density NM EoS following the formalism of Ref.~\cite{Annala:2019puf}. Up to the subnuclear density $\sim 0.5\,n_0$, we use the tabulated BPS EoS~\cite{BPS}. Continuing up to a density of $\sim 1.1\,n_0$, we use polytropes of the form $P = Kn^{\Gamma}$, with $\Gamma \in [1.77, 3.23]$ chosen to span the CET band~\cite{Hebeler_2013}. Above this density, we implement a five-segment speed-of-sound interpolation technique as discussed in \ref{eos_construction}. 
%following Ref.~\cite{Annala2020}. This method parametrises the speed of sound ($c_s^2$) as a function of chemical potential ($\mu$), which allows us to write the number density as 
% \begin{equation}
% n(\mu) = n_{\scriptscriptstyle CET}\exp\left[\int_{\mu_{CET}}^{\mu} \frac{d\mu'}{\mu' c_s^2(\mu')}\right]
% \label{1}
% \end{equation}
% where $n_{CET}$ is fixed by matching the lower-density CET EoS to the initial values of the interpolated EoS, and $\mu_{CET}$ is the chemical potential corresponding to $n_{CET}$. Pressure can then readily be calculated as
% \begin{equation}
% p(\mu) = p_{\scriptscriptstyle CET} \! + n_{\scriptscriptstyle CET}\int_{\mu_{CET}}^{\mu}d\mu' \exp\left[\int_{\mu_{CET}}^{\mu'}\frac{d\mu''}{\mu''c_s^2(\mu'')}\right]
% \label{2}
% \end{equation}   
% where $p_{CET}$ is the pressure at $n_{CET}$. The interpolation uses five randomized segments $(c_{s,i}^2,\mu_{i})$, where each $\mu_{i}$ is randomized within $[\mu_{\scriptscriptstyle CET},\,2.6~\mathrm{GeV}]$. The upper limit of $\mu$ follows Ref.~\cite{Kurkela_2014}, chosen such that the uncertainty in the pQCD regime is roughly the same as the uncertainty at CET.
Finally, the pQCD constraints are enforced as given in \cite{Fraga:2013qra,Altiparmak:2022}. The parameterized pQCD result for cold quark matter in $\beta$ equilibrium is given by:
\begin{equation}
    P_{_{QCD}}(\mu, X) =  \frac{\mu^4}{108\pi^2}\left(c_1 - \frac{d_1X^{-\nu_1}}{\mu/GeV - d_2X^{-\nu_2}}\right) 
\end{equation}
where $c_1 = 0.9008$, $d_1 = 0.5034$, $d_2 = 1.452$, $\nu_1 = 0.3553$, $\nu_2 = 0.9101$, and the renormalization scale parameter $X \in[1,4]$.

\subsubsection{Dark matter EoS}
Various dark matter models, including fermionic and bosonic scenarios, have already been extensively studied \cite{NarainSchaffnerBielich2006,TolosSchaffnerBielich2015,IvanytskyiSagunLopes2020,NelsonReddyZhou2019,Karkevandi2022,Giangrandi2023,PanotopoulosLopes2017,DasMalikNayak2019,DasMalikNayak2022}. In the present work, however, we do not attempt to favor or prioritize any specific microscopic dark matter model. Instead, we adopt an agnostic approach for constructing the dark matter EoS. To maintain consistency with the treatment of nuclear matter, we employ the same speed-of-sound formalism for dark matter as well. 
\textcolor{black}{The choice of the SOSI parametrization for the DM EoS is motivated by our aim to cover as much of the ($P , \epsilon$) parameter space as possible. This allows us to accommodate a broad range of possible DM EoSs without introducing a bias towards any particular microscopic model. As any physically consistent EoS must satisfy the requirements of causality and thermodynamic consistency, a set of SOSI parametrizations can, in principle, cover the entire physically permitted ($P, \epsilon$) space.}

Within this framework, the dark matter EoS is constructed by specifying the dark matter speed-of-sound profile, as given by Eq.~(\ref{3}), together with an anchor point, $(c_{s,\mathrm{start}}^2,\mu_{\mathrm{start}})$. 
% Starting from this anchor point, Eqs.~(\ref{1}) and~(\ref{2}) are integrated up to $\mu_{\mathrm{end}} = 2.6\,\mathrm{GeV}$, thereby obtaining the complete thermodynamic description of the dark matter EoS. 
For simplicity, the low-density regime of dark matter is modeled using a free Fermi gas, parametrized by the so-called dark matter bare mass, $m_D$. We assume that this description remains valid up to densities below $0.1\,n_0$. \textcolor{black}{A detailed discussion of the choice of the low-density anchor at $0.1n_0$, including its implications, is provided in Appendix~\ref{anchor_pnt}.} The thermodynamic quantities evaluated at $0.1\,n_0$ are then used to define the anchor point for the subsequent speed-of-sound based integration at higher densities.

The EoS of fermionic DM at low density can be evaluated as
\begin{equation}\label{eqn5}
\begin{aligned}
\epsilon_{DM} &= \frac{1}{8\pi^2 (\hbar c)^3} \Big[
k_F \sqrt{k_F^2 + m_D^2}\,(2k_F^2 + m_D^2) \\
&\quad - m_D^4 \ln\left(\frac{k_F + \sqrt{k_F^2 + m_D^2}}{m_D}\right)
\Big]
\end{aligned}
\end{equation}

\begin{equation}\label{eqn6}
\begin{aligned}
P_{DM} &=\frac{1}{24\pi^2 (\hbar c)^3}
\Big[k_F \sqrt{k_F^2 + m_D^2}\,(2k_F^2 - 3m_D^2)\\
&\quad + 3m_D^4 \ln\left(\frac{k_F + \sqrt{k_F^2 + m_D^2}}{m_D}\right)\Big]
\end{aligned}
\end{equation}
Here, $k_F$ is the Fermi wave vector of the DM particle, and the number density of DM is given as
\begin{equation}\label{eqn7}
    n_{DM} = \frac{(\hbar k_F)^{3}}{3\pi^2}
\end{equation}

The agnostic speed-of-sound interpolation of the DM EoS is initiated at a number density of \( n = 0.1\,n_0 \). The initial conditions required to perform the integrals in Eqs.~(\ref{1}) and~(\ref{2}) are taken from Eqs.~(\ref{eqn5}), (\ref{eqn6}), and~(\ref{eqn7}), all evaluated at the density(\( n = 0.1\,n_0 \)). The corresponding initial value of the chemical potential is then calculated using \( \mu = \sqrt{k_F^2 + m_D^2} \) at \( n = 0.1\,n_0 \). The interpolation employs five randomized segments, $(c_{s,i}^2,\mu_i)$, where each $\mu_i$ is randomly sampled within the range $[\mu_{\mathrm{start}}$, $2.6\mathrm{GeV}]$. 
% The upper limit of the chemical potential is chosen such that the high-density regime of the DM EoS spans the same range as the NM EoS, enabling a consistent comparison between the two.

\subsection{Dark matter admixed NSs model}
 The structure of a nonrotating, spherically symmetric DMANS, consisting of two perfect fluids---dark matter and nuclear matter ---is described by the modified two-fluid TOV equations~\cite{Tolman1939,OppenheimerVolkoff1939}. With the two components interacting solely through gravity, these read:
\begin{align}
\frac{dP_{\mathrm{NM}}}{dr} &= -\frac{(P_{\mathrm{NM}} + \varepsilon_{\mathrm{NM}})\left\{m + 4\pi r^3(P_{\mathrm{NM}} + P_{\mathrm{DM}})\right\}}{r(r - 2m)} \label{2tov.psr1}\\
\frac{dP_{\mathrm{DM}}}{dr} &= -\frac{(P_{\mathrm{DM}} + \varepsilon_{\mathrm{DM}})\left\{m + 4\pi r^3(P_{\mathrm{NM}} + P_{\mathrm{DM}})\right\}}{r(r - 2m)} \label{2tov.psr2}
\end{align}

with
\begin{IEEEeqnarray}{rCl}
\frac{dm(r)}{dr} &=& 4\pi (\varepsilon_{\mathrm{NM}} + \varepsilon_{\mathrm{DM}})r^2 \label{2tov.mass}
\end{IEEEeqnarray}
Here, $\varepsilon_{\mathrm{NM}}$ and $P_{\mathrm{NM}}$ denote the energy density and pressure of NM, respectively, while $\varepsilon_{\mathrm{DM}}$ and $P_{\mathrm{DM}}$ represent the corresponding quantities for DM. The total gravitational mass enclosed within a radius $r$ is given by $m(r) = m_{\mathrm{NM}}(r) + m_{\mathrm{DM}}(r)$, where $m_{\mathrm{NM}}(r)$ and $m_{\mathrm{DM}}(r)$ are the mass contributions from NM and DM, respectively.

The behavior of the metric potentials inside a DMANS is governed by
\begin{IEEEeqnarray}{rCl}
    \frac{d\nu(r)}{dr} &=& \frac{\left\{m + 4\pi r^3(P_{\mathrm{NM}} + P_{\mathrm{DM}})\right\}}{r(r - 2m)}\label{2tov.nu}\\
    \lambda(r) &=& -\frac{1}{2}\ln{\left\{1 - \frac{2(m_{\mathrm{NM}} + m_{\mathrm{DM}})}{r}\right\}}\label{2tov.lamda}
\end{IEEEeqnarray}

The coupled two-fluid TOV equations~(\ref{2tov.psr1}), (\ref{2tov.psr2}), and~(\ref{2tov.mass}), together with the metric-potential equations~(\ref{2tov.nu}) and~(\ref{2tov.lamda}), can be solved for given EoSs of DM and NM, subject to appropriate boundary conditions. The surfaces of the DM and NM components are defined by the vanishing of their respective pressures, i.e., $P_{\mathrm{DM}}(R_{\mathrm{DM}}) = 0, \quad P_{\mathrm{NM}}(R_{\mathrm{NM}}) = 0,$ where $R_{\mathrm{DM}}$ and $R_{\mathrm{NM}}$ denote the radii of the DM and NM components, respectively. The actual radius of the star is taken to be $R = \max\left(R_{\mathrm{NM}}, R_{\mathrm{DM}}\right)$,
whereas the observable radius corresponds to the NM radius, $R_{\mathrm{NM}}$.

The amount of DM contained within a neutron star is quantified by the DM fraction, defined as
\begin{IEEEeqnarray}{rCl}
    f_{\mathrm{DM}} &=& \frac{M_{\mathrm{DM}}}{M_{\mathrm{NM}} + M_{\mathrm{DM}}},
\end{IEEEeqnarray}
where $M_{\mathrm{DM}}$ and $M_{\mathrm{NM}}$ are the total masses of the DM and NM components, respectively. The DM content in a DMANS can thus be controlled by varying this fraction.

The tidal response of a spherically symmetric DMANS in a static external quadrupolar tidal field is quantified by the dimensionless tidal deformability $\Lambda$, defined as
\begin{IEEEeqnarray}{rcl}
    \Lambda = \frac{2}{3}k_{2} \left(\frac{R}{M}\right)^{5}.
    \label{tdf_eq}
\end{IEEEeqnarray}
Here, $M$ and $R$ are the mass and radius of the DMANS, respectively, and $k_{2}$ is the dimensionless tidal Love number~\cite{Hinderer2008,DamourNagar2009,Hinderer2010}, which depends on the internal structure of the star and is defined as
\begin{IEEEeqnarray}{rcl}
k_2 &=& \frac{8 C^5}{5} (1 - 2C)^2 \left[ 2 + 2C \left(y(R) - 1\right) - y(R) \right] \nonumber \\
&& \times \Bigg\{
2C \left( 6 - 3y_R + 3C\left(5y(R) - 8\right) \right) \nonumber \\
&& \quad + 4C^3 \left[ 13 - 11y(R) + C\left(3y(R) - 2\right) + 2C^2 \left(1 + y(R)\right) \right] \nonumber \\
&& \quad + 3(1 - 2C)^2 \left[ 2 - y(R) + 2C\left(y(R) - 1\right) \right] \nonumber \\
&& \times \log(1 - 2C)
\Bigg\}^{-1}
\end{IEEEeqnarray}
Here, \( C = M/R \) denotes the compactness of the DMANS, and \( y(R) \) is obtained by solving the following differential equation~\cite{Postnikov2010} from the center to the stellar surface, along with the two-fluid TOV equations, and subject to appropriate boundary conditions.
\begin{IEEEeqnarray}{rcl}
r \frac{dy(r)}{dr} &+&  \, y(r)^2 \nonumber \\
&& +\, y(r)\, e^{2\lambda} \left(1 + 4\pi r^2 (P - \varepsilon)\right) \nonumber \\
&& +\, r^2 e^{2\lambda} \left(4\pi \left[ 5\varepsilon + 9P + \frac{(P + \varepsilon)}{C_{\mathrm{eff}}^{2}} \right] - \frac{6}{r^2} \right) \nonumber \\
&& -\, 4r^{2} \left( \frac{d\nu}{dr} \right)^2 = 0
\end{IEEEeqnarray}
Where \( P = P_{NM} + P_{DM} \) is the total pressure, \( \varepsilon = \varepsilon_{NM} + \varepsilon_{DM} \) is the total energy density of the DMANS, and \( C_{\mathrm{eff}}^{2} \) is the effective speed of sound, defined as 
\begin{equation}
\left(C_{\mathrm{eff}}^{2}\right)^{-1} 
= \left[ \frac{\varepsilon_{NM} + P_{NM}}{C_{s,NM}^{2}} 
+ \frac{\varepsilon_{DM} + P_{DM}}{C_{s,DM}^{2}} \right] 
\frac{1}{P + \varepsilon}
\end{equation}
where $C_{s,NM}^{2}$ and $C_{s,DM}^{2}$ are the squared speeds of sound of NM and DM, respectively.

\section{Results}

 In the following, we present the astrophysical constraints that are imposed to obtain the main results of our study. The next subsections discuss the effects of DM on the EoS, the NS mass and radius, mass and tidal deformability, and possible configurations of DM distribution in halo-like or core-like settings. \textcolor{black}{The main assumption of our SOSI agnostic approach is (a) the DM is assumed to be a free fermionic DM till the anchor density, (b) the DM and NM interact among themselves only through gravity.}

\subsection{Astrophysical Constraints}
\label{astro_const}
To systematically investigate the impact of current astrophysical observations on the properties of dark matter-admixed neutron stars, we impose four sets of observational constraints on the stellar configurations, defined as follows:

\begin{enumerate}[label=(\roman*)]

\item \label{cons1} NICER $1\sigma$: The mass-radius constraints derived from NICER X-ray timing observations of four millisecond pulsars---PSR~J0614$-$3329~\cite{Mauviard2025}, PSR~J0740$+$6620~\cite{Riley2021,Miller2021,Salmi2024,Dittmann2024}, PSR~J0030$+$0451~\cite{Riley2019,Miller2019}, and PSR~J0437$-$4715~\cite{Choudhury2024}---  combined with the mass-radius posterior from GW170817~\cite{GW170817}, within the $1\sigma$ confidence interval.

\item \label{cons2}  NICER $2\sigma$: Same as above, but extended to the $2\sigma$ confidence interval, thereby covering a broader range of permissible mass-radius configurations.

\item \label{cons3}  GW+2.01: The tidal deformability constraint from the binary neutron star merger event GW170817, which restricts the dimensionless tidal deformability of a canonical $1.4\,M_\odot$ star to the range $70 \leq \Lambda_{1.4} \leq 580$~\cite{GW170817}, combined with the maximum mass requirement $M_{\mathrm{max}} \geq 2.01\,M_\odot$~\cite{Antoniadis2013}.

\item \label{cons4}  GW+2.01+NICER $1\sigma$: The tightest combined constraint, requiring simultaneous satisfaction of the NICER $1\sigma$ mass-radius constraint and the GW+2.01 tidal deformability and maximum mass constraints.

\end{enumerate}

These four constraint sets are applied uniformly across all DM models considered in this work, enabling a systematic and comparative statistical analysis of the allowed DMANS parameter space. As may be seen, the uncertainties considered are almost always at the level of 1$\sigma$, which may be considered to be limitative. This choice allows us to understand whether future observations, which are expected to come with smaller uncertainties, will be significant. The effects of the broader astrophysical constraints are also discussed in Appendix~\ref{broad_apc}.

\subsection{Equations of state}
\begin{figure}[ht]
    \centering
    \includegraphics[width=0.45\textwidth]{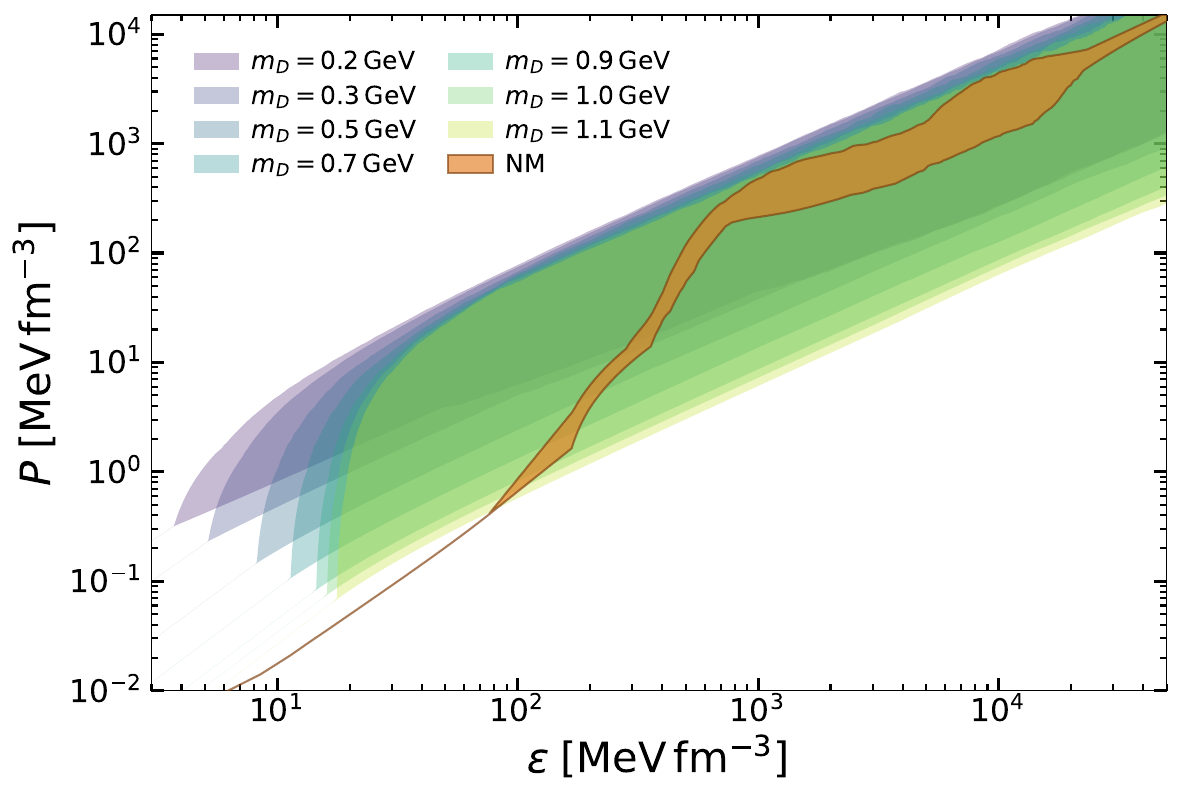}
    \caption{Equations of state (pressure $P$ as a function of energy density $\varepsilon$) for dark matter models with different particle masses, $m_D$ (shown in different colors), together with the nuclear matter (NM) equation of state (orange), extended to pQCD densities. All curves are plotted on logarithmic scales.}
\label{eos}
\end{figure}

In this work, we adopt an agnostic framework for both the NM EoS and DM EoS. To construct DMANS configurations, we first select NM EoSs that satisfy current observational constraints, including the mass-radius measurements reported by NICER and those inferred from the gravitational-wave event GW170817, particularly the constraints on tidal deformability. The NM EoSs shown as the orange contour in Fig.~\ref{eos} represent the subset of well-constrained models employed in this study, which lie within the $1\sigma$ confidence region of NICER observations and are consistent with the tidal deformability constraints from GW170817, constraints \ref{cons4} defined above. Furthermore, these EoSs are smoothly extended to the pQCD regime at high densities, ensuring consistency with fundamental nuclear physics and theoretical constraints.

For the DM sector, we also adopt an agnostic approach by employing a speed-of-sound parametrization with randomized sampling. Using this framework,  we construct seven distinct dark matter EoS models corresponding to particle masses in the range $m_D = 0.2$--$1.1\,\mathrm{GeV}$, spanning values below and above the nucleon mass. \textcolor{black}{The mass range is consistent with other works that exist previously in the literature \cite{Kain,Barbat:2024yvi,Mariani:2023wtv}}.
This choice enables a systematic investigation of halo-like and core-like dark matter configurations. These models are represented by differently colored contours in Fig.~\ref{eos}. As expected, the smaller the mass of the dark particle, the larger the pressure for a given energy density.

\subsection{Mass and radius}
The equilibrium structure of DMANS is obtained by solving the two-fluid TOV equations (Eqs.~\ref{2tov.psr1}, \ref{2tov.psr2}, and \ref{2tov.mass}) for pairs of NM and DM equations of state, along with the dark matter fraction $f_{\mathrm{DM}}$. These EoSs are randomly sampled to ensure fair coverage of the model space considered in this study. For each configuration, $f_{\mathrm{DM}}$ is randomly selected within the range $0.01 \leq f_{\mathrm{DM}} \leq 0.15$. This procedure enables us to generate an ensemble of approximately $10^{5}$ mass-radius (M--R) sequences for DMANS corresponding to each dark matter model, thereby enabling a systematic exploration of the influence of dark matter properties on neutron star structure.

\begin{figure*}[ht]
    \centering
    \includegraphics[width=0.45\textwidth]{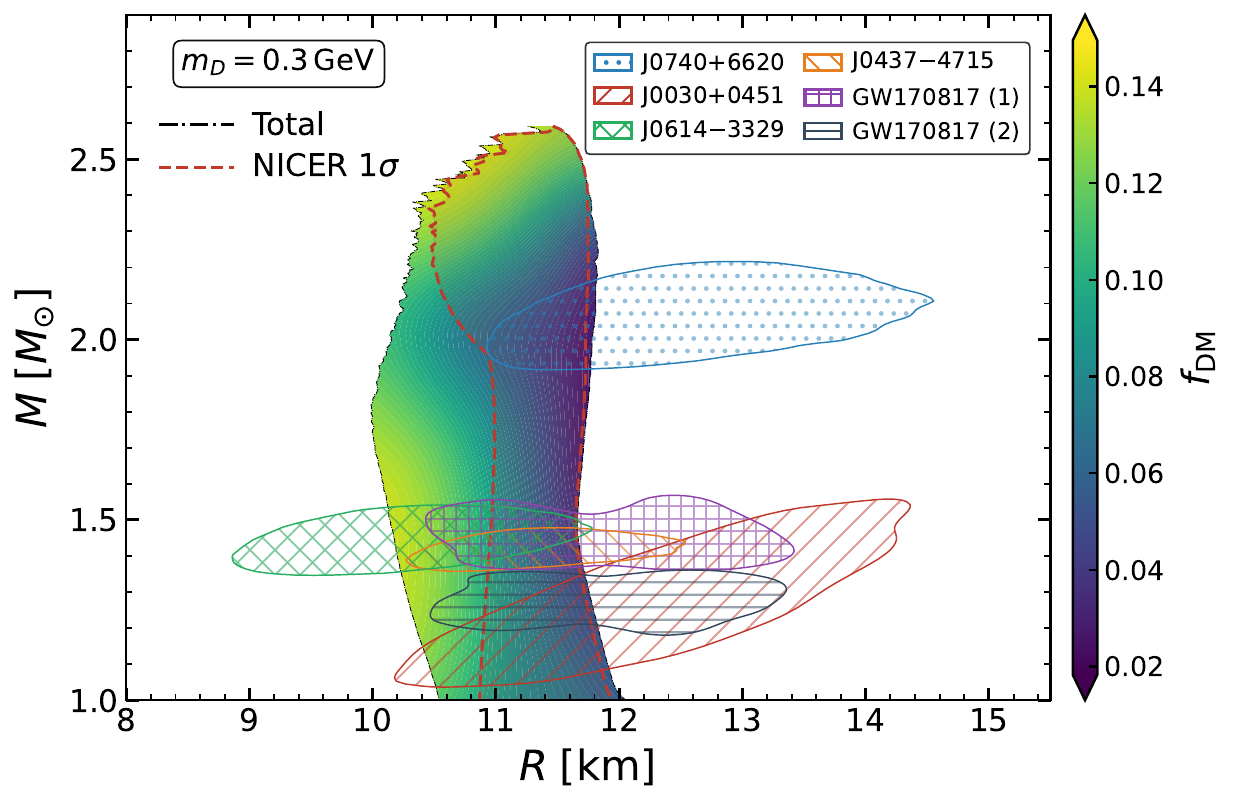}\hfill
    \includegraphics[width=0.45\textwidth]{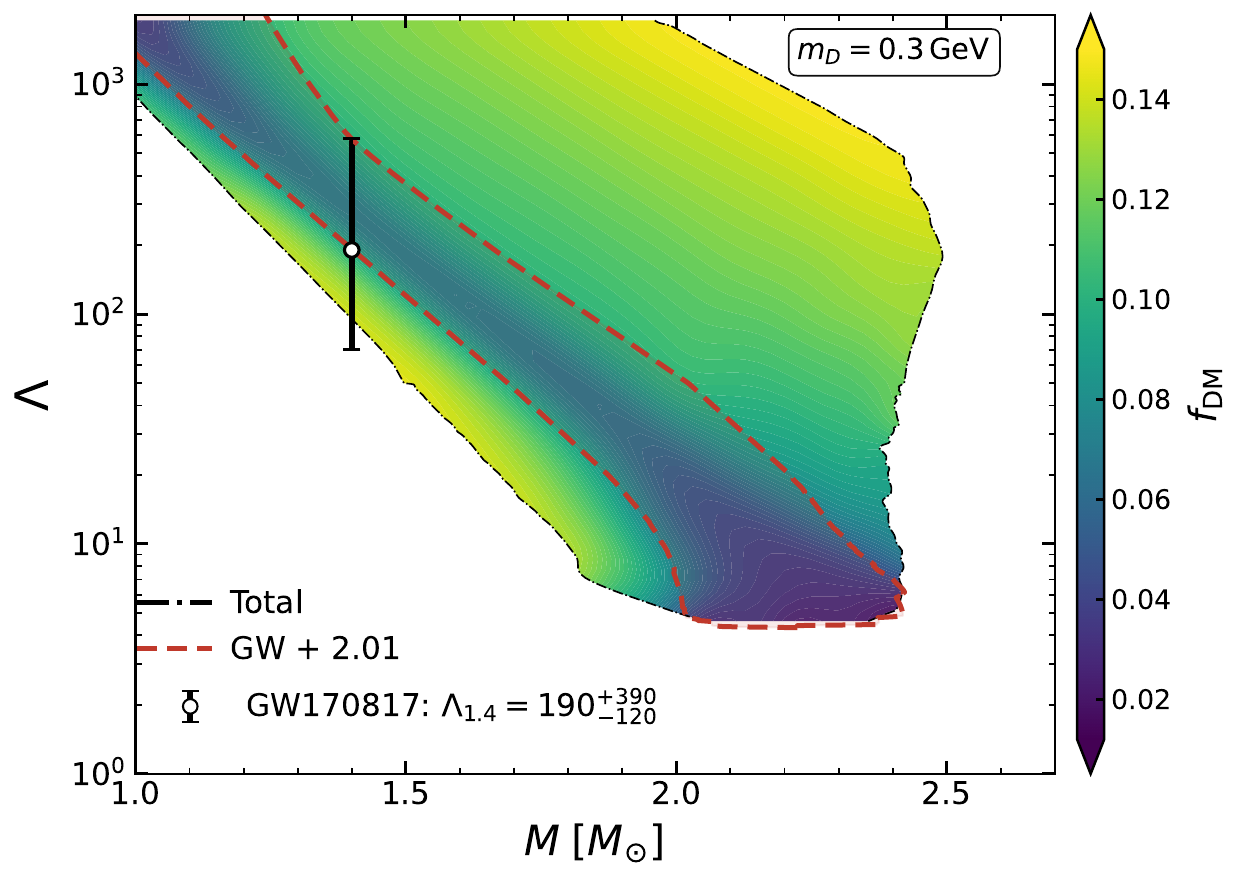}\\
    \includegraphics[width=0.45\textwidth]{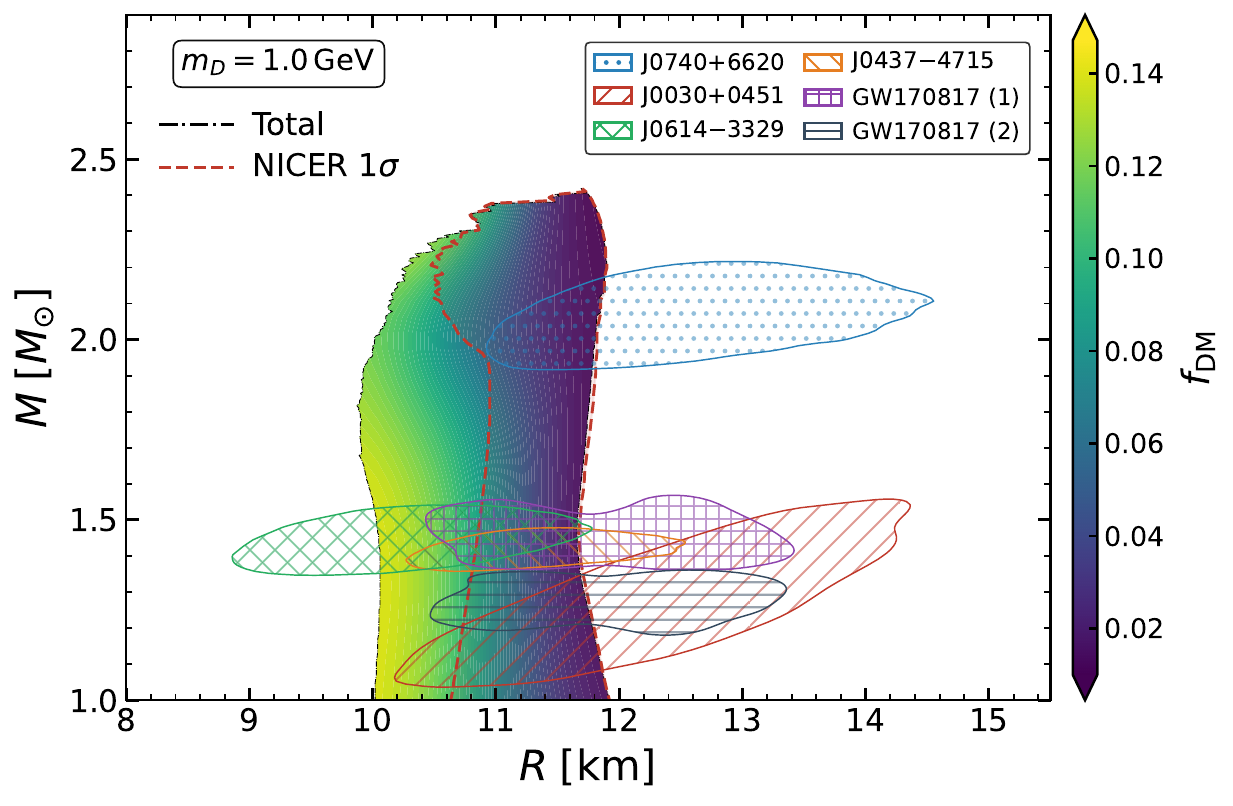}\hfill
    \includegraphics[width=0.45\textwidth]{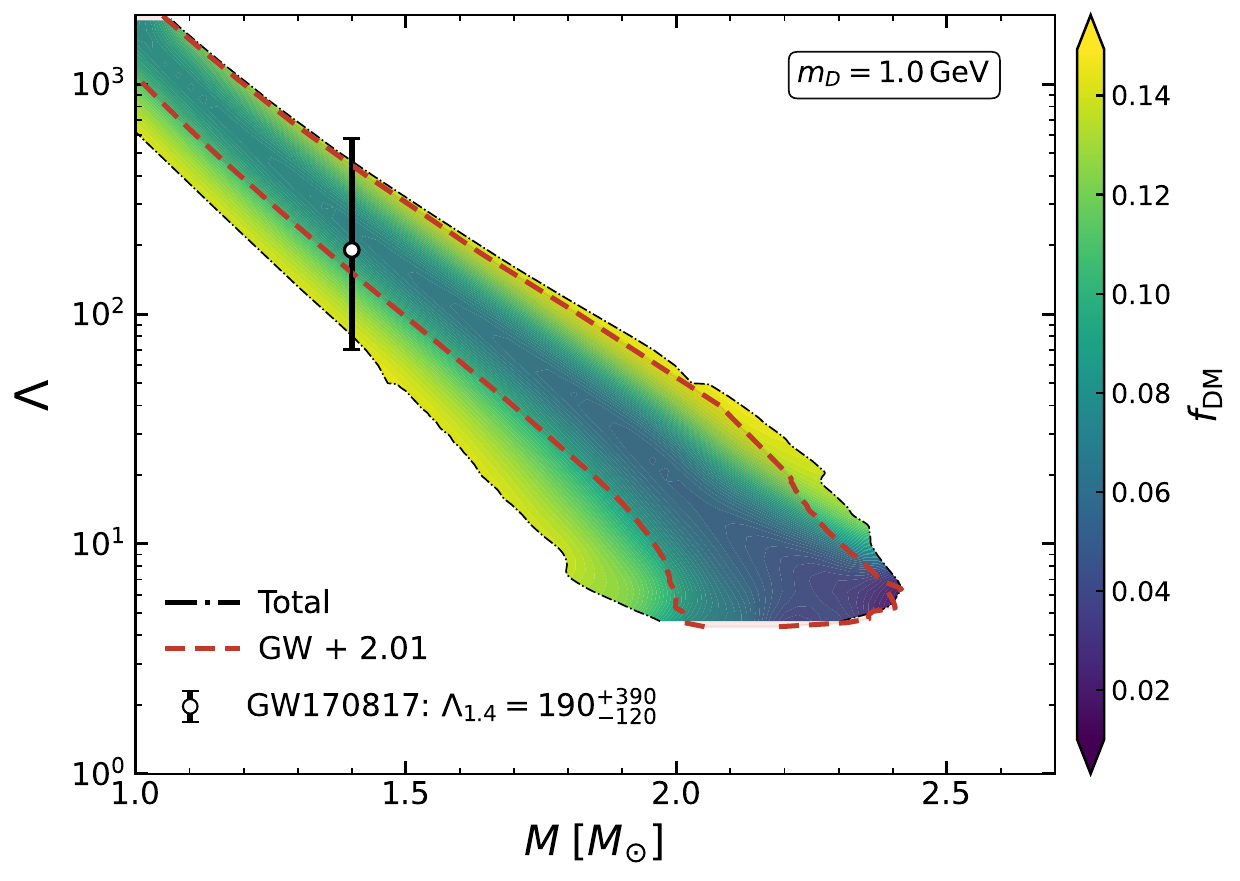}
    \caption{Mass-radius (M--R), with $R=R_{NM}$, relations (\textbf{left column}) and dimensionless tidal deformability $\Lambda$ as a function of stellar mass (\textbf{right column}) for DMANS, shown for two dark matter models: $m_D = 0.3\,\mathrm{GeV}$ (\textbf{top row}) and $m_D = 1.0\,\mathrm{GeV}$ (\textbf{bottom row}). In every panel, the color gradient encodes the dark matter fraction $f_{\mathrm{DM}}$. In the $M-R_{NM}$ panels (left) the black dash-dotted contour encloses the complete set of theoretically allowed sequences, while the red dashed contour marks those satisfying the combined NICER and GW170817 mass-radius constraints at the $1\sigma$ level; the remaining marked regions correspond to the individual NICER and GW170817 constraints, as labeled. In the $\Lambda$--$M$ panels (right) the black dash-dotted contour encloses all theoretically allowed tidal responses, while the red dashed contour identifies the subset consistent with the GW170817 bound $70 \leq \Lambda_{1.4} \leq 580$ (indicated by the black horizontal bar).}
    \label{mr_tdf}
\end{figure*}

The left column of Fig.~\ref{mr_tdf} shows the M--R sequences for DMANS for two representative dark matter models, $m_D = 0.3\,\mathrm{GeV}$ (top) and $m_D = 1.0\,\mathrm{GeV}$ (bottom); we first discuss the $m_D = 0.3\,\mathrm{GeV}$ case in detail. From the electromagnetic observations, it is possible to infer the nuclear matter radius $R_{NM}$. Therefore, the stellar radius measured coincides with the nuclear matter radius, $R \equiv R_{\mathrm{NM}}$. The color gradient represents the dark matter fraction $f_{\mathrm{DM}}$, spanning from $f_{\mathrm{DM}} = 0.01$ (low, shown in purple) to $f_{\mathrm{DM}} = 0.15$ (high, shown in yellow), thereby illustrating the systematic impact of increasing dark matter content on the stellar structure. It is evident that as $f_{\mathrm{DM}}$ increases, the M--R sequences are systematically affected in two distinct ways, depending on the underlying DM equation of state. In the low to intermediate mass regime, increasing $f_{\mathrm{DM}}$ shifts the M--R sequences progressively toward smaller radii at a given mass, indicating that the accumulation of dark matter softens the effective EoS and renders the star more compact. However, in the higher-mass regime---as evidenced by the yellow-colored band at the upper end of the M--R sequences---a higher dark matter fraction can also support larger maximum masses, suggesting that the dark matter component, depending on its EoS, may provide additional pressure support that effectively stiffens the total EoS at higher central densities. This dual behavior highlights the nontrivial role of dark matter in neutron star structure: while it compactifies the star at moderate masses, it can simultaneously extend the maximum mass limit under certain dark matter EoS conditions.

\begin{figure*}[ht]
    \centering
    \includegraphics[width=0.99\textwidth]{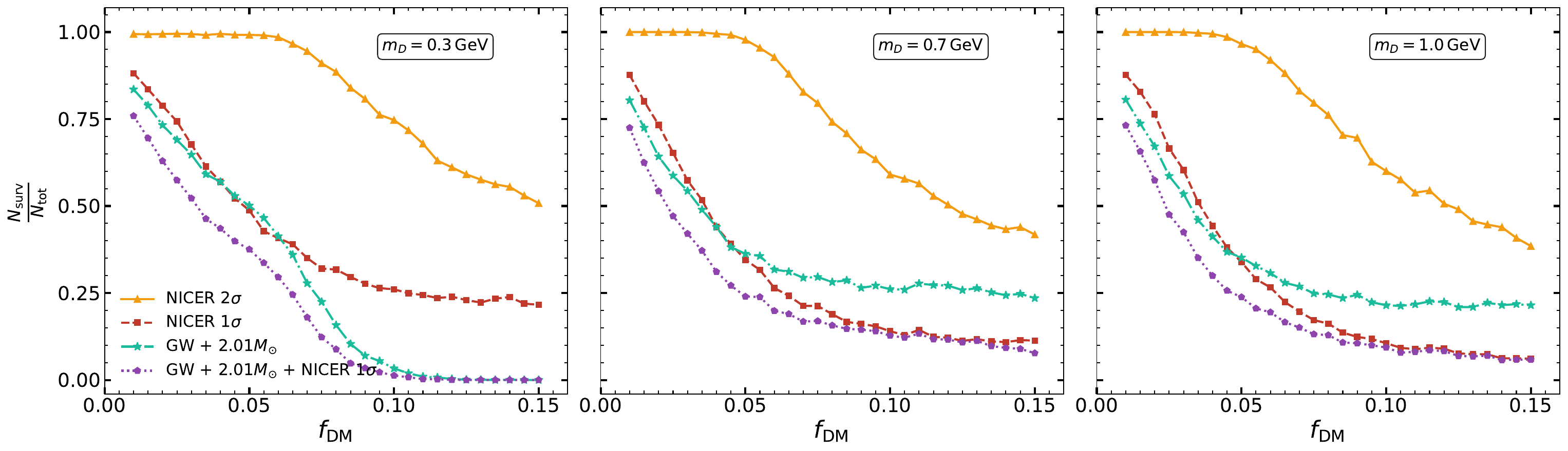}
    \caption{ normalized survival fraction of M--R sequences, $N_{\mathrm{surv}}/N_{\mathrm{tot}}$, as a function of the dark matter fraction $f_{\mathrm{DM}}$, shown for three dark matter models: (\textbf{Left}) $m_D = 0.3\,\mathrm{GeV}$, (\textbf{Middle}) $m_D = 0.7\,\mathrm{GeV}$, and (\textbf{Right}) $m_D = 1.0\,\mathrm{GeV}$. In each panel, the four curves correspond to the observational constraint sets defined in the text~\ref{astro_const}: NICER $2\sigma$ (orange, triangles), NICER $1\sigma$ (red, squares), GW+2.01 (cyan, stars), and the combined GW+2.01+NICER $1\sigma$ (purple, pentagons).}
 \label{norm_all}
\end{figure*}

The outer black dash-dotted contour, labeled ``Total'', spans the complete set of theoretically allowed M--R sequences for DMANS with dark matter fractions up to $f_{\mathrm{DM}} \leq 15\%$, representing the full parameter space explored in this work. The inner red dashed contour, labeled ``NICER $1\sigma$'', outlines the observationally viable subset of these M--R sequences, i.e., those that simultaneously satisfy the mass-radius constraints from NICER observations of PSR~J0614$-$3329~\cite{Mauviard2025}, PSR~J0740$+$6620~\cite{Riley2021,Miller2021,Salmi2024,Dittmann2024}, PSR~J0030$+$0451~\cite{Riley2019,Miller2019}, and PSR~J0437$-$4715~\cite{Choudhury2024}, as well as the MR constraints derived from the gravitational wave event GW170817~\cite{GW170817}, all within the $1\sigma$ confidence interval.

A comparison of the two contours reveals an important observational constraint on the dark matter content of neutron stars. Specifically, M--R configurations characterized by high dark matter fractions and relatively low stellar masses are excluded by the current NICER and GW170817 constraints, as these configurations produce stars that are too compact to be consistent with the observed M--R windows. This indicates that the current observational data disfavor highly dark matter-admixed configurations at lower stellar masses. However, note that these low mass-low radius configurations are compatible with HESS  J1731-347 \cite{Hess2022}.

The bottom-left panel of Fig.~\ref{mr_tdf} presents the M--R sequences for the dark matter model with $m_D = 1.0~\mathrm{GeV}$. The color scale denotes the dark matter distribution for different stellar configurations. An increase in the dark matter content results in the formation of a denser dark matter core, making the star more compact and shifting the corresponding M--R sequence toward smaller radii. Comparison with the NICER mass-radius constraints reveals that configurations containing a larger fraction of dark matter are largely excluded, while those with lower dark matter content remain consistent with the observational data

Overall, the comparison between the two representative dark matter models reveals a strong dependence of the NS structure on the DM particle mass. For the lower-mass DM model ($m_D = 0.3\mathrm{GeV}$), the DM component can form both extended halo-like distributions and dense cores, leading to a rich variety of stellar configurations. In this case, the presence of DM not only increases the compactness of the star but can also support configurations with larger masses and radii, depending on the dark matter fraction and the underlying dark matter EoS. In contrast, for the higher-mass dark matter model ($m_D = 1.0\,\mathrm{GeV}$), the dark matter is predominantly concentrated in a dense central core, resulting primarily in a more compact stellar structure with reduced radii. Consequently, while low-mass dark matter can generate both halo and core structures and significantly modify the global mass-radius properties of the star, high-mass dark matter mainly acts to compactify the star through the formation of a central dark matter core.

\subsection{Tidal deformability}
To investigate the impact of tidal deformation on DMANS, we solve Eq.~(\ref{tdf_eq}) and present in the right column of Fig.~\ref{mr_tdf} the dimensionless tidal deformability, $\Lambda$, as a function of the stellar mass, $M$, for the $m_D = 0.3~\mathrm{GeV}$ (top) and $m_D = 1.0~\mathrm{GeV}$ (bottom) models; the discussion below focuses on the $m_D = 0.3~\mathrm{GeV}$ case. The color gradient represents the dark matter fraction, $f_{\mathrm{DM}}$, while the outer dash-dotted contour encloses the full set of theoretically allowed $\Lambda$--$M$ configurations for DMANS across the explored range of $f_{\mathrm{DM}}$.

The distribution of $f_{\mathrm{DM}}$ across the $\Lambda$-$M$ plane exhibits a nontrivial dependence of tidal deformability on dark matter content. Since tidal deformability scales sensitively with the stellar radius as $\Lambda \propto R^5$  ($R=$Max$(R_{NM},R_{DM}$)), neutron stars containing extended dark matter halos characterized by a larger effective radius tend to exhibit enhanced tidal deformability at a fixed mass compared to purely hadronic stars. This behavior is evident in the upper region of the parameter space, where configurations with higher $f_{\mathrm{DM}}$ and extended halos correspond to larger values of $\Lambda$.

In contrast, in the low-mass regime, configurations with large $f_{\mathrm{DM}}$ that develop compact dark matter cores, rather than extended halos, result in a more compact stellar structure and consequently lower tidal deformability. This highlights the dual role of dark matter -- core-dominated versus halo-dominated -- in governing the tidal response of DMANS.

The inner red dashed contour, labeled ``GW+2.01,'' denotes the subset of configurations that simultaneously satisfy two key observational constraints: (i) the tidal deformability bound from the binary neutron star merger GW170817, which constrains the deformability of a $1.4\,M_\odot$ star to $70 \leq \Lambda_{1.4} \leq 580$~\cite{GW170817}, and (ii) the maximum mass requirement $M_{\mathrm{max}} \geq 2.01\,M_\odot$~\cite{Antoniadis2013}. The imposition of these combined constraints significantly reduces the allowed parameter space. In particular, configurations with large dark matter fractions, especially those in the halo-dominated regime, are strongly disfavored, as they predict tidal deformabilities exceeding the upper bound inferred from GW170817.

The lower panel of Fig.~\ref{mr_tdf} presents the tidal response of the $m_D = 1.0 \mathrm{GeV}$ dark matter model. In contrast to the $m_D = 0.3 \mathrm{GeV}$ case, the stellar configurations in this model are predominantly core-dominated, with the dark matter concentrated in a dense central region rather than forming extended halo structures. The resulting increase in stellar compactness leads to systematically smaller tidal deformabilities at a given mass. Consequently, the majority of configurations naturally satisfy the tidal deformability constraints inferred from GW170817. Moreover, some configurations with larger mass and relatively large dark matter fractions can still satisfy the combined observational constraint ``GW+2.01''.

Overall, our results demonstrate that the tidal response of DMANSs is highly sensitive to both the dark matter fraction and its spatial distribution. While halo-dominated configurations tend to produce enhanced tidal deformabilities and are therefore strongly constrained by GW170817, core-dominated configurations remain largely compatible with current observations owing to their increased compactness. The combined tidal deformability and maximum-mass constraints thus provide a robust framework for probing the presence of dark matter in neutron stars and for placing stringent limits on the properties of dark-matter-admixed configurations.

\subsection{Distribution of Permissible Dark Matter Fractions Under
Astrophysical Constraints}

\label{fdm_const}
\begin{figure*}[ht]
    \centering
    \includegraphics[width=0.99\textwidth]{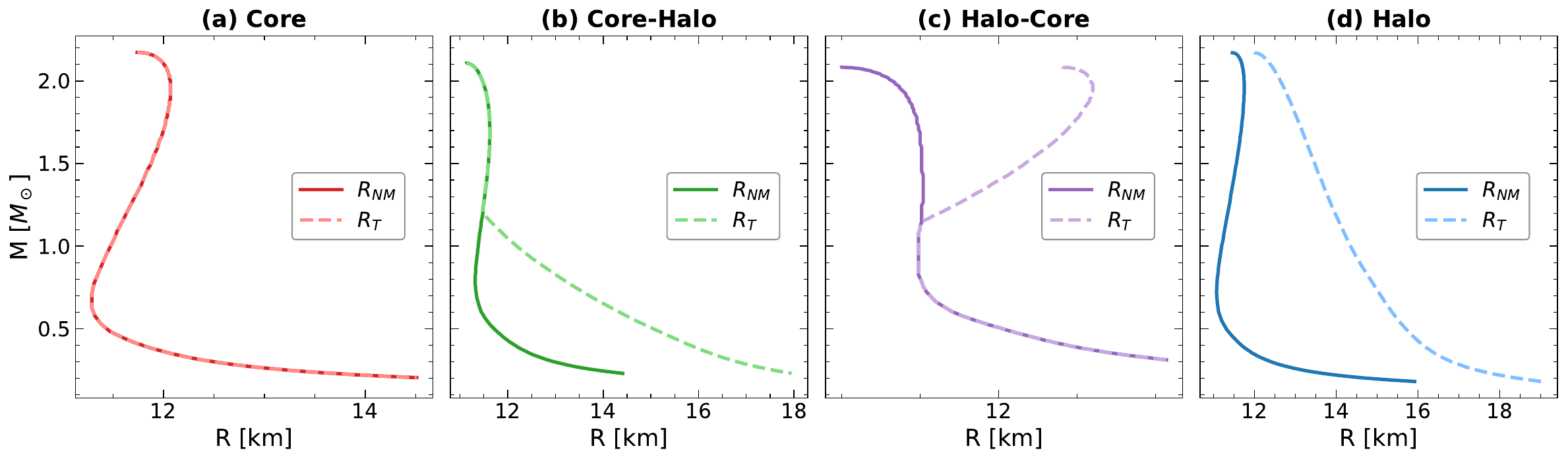}
    \caption{ Four possible mass-radius configuration types for DMANS  as a function of $R=$Max$(R_{NM}, R_{DM})$: (a) Core, where all stars possess dark matter cores ($R = R_{\mathrm{NM}}$); (b) Core-Halo, where lower-mass stars exhibit dark matter halos ($R_{\mathrm{NM}} < R_{\mathrm{DM}}$, so $R = R_{\mathrm{DM}}$ )  and higher-mass stars contain dark matter cores ($R= R_{\mathrm{NM}}$); (c) Halo-Core, where higher-mass stars exhibit dark matter halos ($R_{\mathrm{NM}} < R_{DM}$) so $R = R_{\mathrm{DM}}$ and lower-mass stars contain dark matter cores ($R = R_{\mathrm{NM}}$); and (d) Halo, where all stars possess extended dark matter halos ($R = R_{\mathrm{DM}}$). The solid line represents the nucleonic matter radius, $R_{\mathrm{NM}}$, while the dashed line denotes the total radius of the DMANS, $R_{\mathrm{T}}$. For configurations with an extended dark matter halo ($R_{\mathrm{T}} \neq R_{\mathrm{NM}}$), the total radius is determined by the dark matter distribution, i.e., $R_{\mathrm{T}} \equiv R_{\mathrm{DM}}$.}
\label{prof_mr}
\end{figure*}

To assess the robustness and generality of the above results, we extend our analysis to additional dark matter models with different dark matter particle masses, $m_D = 0.2,\ 0.5,\ 0.7,\ 0.9,$ and $1.1~\mathrm{GeV}$. Across all cases, the qualitative features observed for $m_D = 0.3~\mathrm{GeV}$   and $m_D = 1.0~\mathrm{GeV}$, including the mass-radius compactification and the behavior of the tidal deformability, remain preserved. Quantitative differences arise among different DM models with varying $m_D$ values, driven by the corresponding variations in the DM EoS. A comprehensive statistical analysis of these models is presented in the following section.

The application of the  four constraint sets \ref{cons1},\ref{cons2}, \ref{cons3},\ref{cons4} to all DM models is summarized in 
Fig.~\ref{norm_all}  through the \emph{normalized survival fraction}, $N_{\mathrm{surv}}/N_{\mathrm{tot}}$, defined at each $f_{\mathrm{DM}}$ as the ratio of the number of M-R sequences that survive a given constraint set to the total number sampled. The three panels correspond to the dark matter models with $m_D = 0.3$, $0.7$, and $1.0~\mathrm{GeV}$, and within each panel, the four curves represent the four constraint sets defined above \ref{astro_const}. Because the unconstrained sample is approximately uniform in $f_{\mathrm{DM}}$, this ratio cleanly isolates the constraining power of each observation as a function of the dark matter content.

For the light mass model, $m_D = 0.3~\mathrm{GeV}$ (left panel), the NICER $2\sigma$ \ref{cons2} constraint provides the weakest filtering and retains a substantial fraction of sequences across the entire $f_{\mathrm{DM}}$ range, while the NICER $1\sigma$\ref{cons1} constraint is systematically more restrictive. The combined GW+2.01 \ref{cons3} constraint is, however, by far the most effective of the individual cuts: the survival fraction falls rapidly with increasing $f_{\mathrm{DM}}$ and approaches zero beyond $f_{\mathrm{DM}} \sim 0.11$, effectively imposing an upper bound on the permissible dark matter fraction. This behavior is consistent with Fig.~\ref{mr_tdf}: for this light mass model, the dark matter tends to form extended halos that inflate the tidal deformability beyond the GW170817 bound, so the tidal measurement is more constraining than the mass-radius data. The combined GW+2.01+NICER $1\sigma$\ref{cons4} cut naturally yields the most stringent restriction of all.

For the heavier models, $m_D = 0.7~\mathrm{GeV}$ (middle panel) and $m_D = 1.0~\mathrm{GeV}$ (right panel), this hierarchy is reversed. Here, the dark matter is increasingly core-dominated and acts to compactify the star, shifting the M--R sequences toward smaller radii; such configurations are more readily excluded by the NICER mass-radius windows than by the tidal-deformability bound, which is most sensitive to extended halos. Consequently, the NICER $1\sigma$ curve falls below the GW+2.01 curve over much of the $f_{\mathrm{DM}}$ range, and the mass-radius data become the dominant constraint on the dark matter fraction.

Taken together, the three panels reveal a clear and physically intuitive trend: as the dark matter particle mass increases and its distribution evolves from halo-like to core-like, the dominant constraint on the dark matter content shifts from the gravitational-wave tidal deformability to the NICER mass-radius measurements.

\subsection{Core and Halo Configurations}
\label{core_halo_config}
\begin{figure*}
    \centering
    \includegraphics[width=0.99\textwidth]{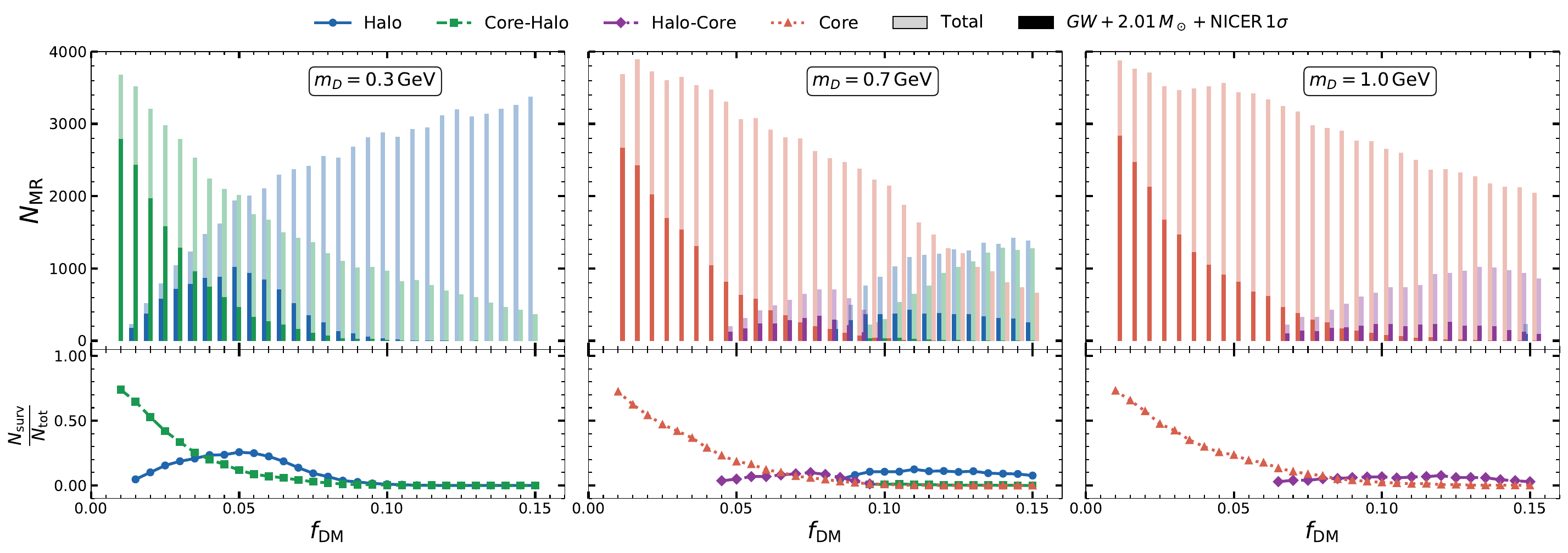}
    \caption{Statistical distribution of the number of M--R relations belonging to four different configuration categories: \textit{Halo} (blue), \textit{Core-Halo} (green), \textit{Core} (orange), and \textit{Halo-Core} (purple), across different values of $f_{\mathrm{DM}}$. (\textbf{Left panel}) Analysis for the dark matter model with $m_D = 0.3\,\mathrm{GeV}$, where the configurations are dominated by \textit{Core-Halo}  and \textit{Halo} structures. The light-colored bars represent the total number of possible M--R relations, while the dark-colored bars indicate the surviving configurations after imposing the most stringent constraint, GW+2.01+NICER $1\sigma$. The lower panel shows the corresponding normalized survival fraction for the above configurations. (\textbf{Middle panel}) Distribution for the dark matter model with $m_D = 0.7\,\mathrm{GeV}$, where the parameter space is dominated by \textit{Core} configurations, along with \textit{Halo-Core} configurations at intermediate dark matter fractions, and smaller contributions from \textit{Core-Halo} and \textit{Halo} (blue) configurations. (\textbf{Right panel}) Same analysis as the left and middle panels, but for the dark matter model with $m_D = 1.0\,\mathrm{GeV}$.}
    \label{halo_all}
\end{figure*}

A dark matter-admixed neutron star can be classified into four distinct categories based on the spatial distribution of dark matter. A configuration is said to have a dark matter core when the nucleonic matter radius exceeds the dark matter radius, i.e., $R_{\mathrm{NM}} > R_{\mathrm{DM}}$, such that the dark matter is confined within the stellar interior, leading to a more compact structure. In contrast, an extended dark matter halo is present when $R_{\mathrm{DM}} > R_{\mathrm{NM}}$, with dark matter distributed outside the nucleonic component.

Depending on the dark matter equation of state, the entire M-R sequence can exhibit different structural characteristics. When all stellar configurations in the sequence possess a dark matter core, the sequence is classified as \textit{Core}. If the sequence contains core-dominated configurations at higher masses and halo-dominated configurations at lower masses, it is denoted as a \textit{Core-Halo} configuration. Conversely, when halo structures appear at higher masses and core structures at lower masses, the sequence is classified as \textit{Halo-Core}. Finally, if all configurations in the sequence exhibit an extended dark matter halo, the sequence is classified as \textit{Halo}. The four types of configurations are illustrated in Fig.~\ref{prof_mr}: (a) \textit{Core}, (b) \textit{Core-Halo}, (c) \textit{Halo-Core}, and (d) \textit{Halo}.

Fig.~\ref{halo_all} presents the statistical distribution of the four 
structural configurations: \textit{Core}, 
\textit{Core-Halo}, \textit{Halo-Core}, and 
\textit{Halo}, as a function of $f_\mathrm{DM}$
for different dark matter models ($m_D = 0.3$, $0.7$, and $1.0\,\mathrm{GeV}$) in the form of bar plots.
For the dark matter model with $m_D = 0.3\,\mathrm{GeV}$, the complete set of DM equations of state exclusively produces 
\textit{Core-Halo} and \textit{Halo} 
configurations, indicating that this model is predominantly 
halo-dominated. Specifically, the \textit{Halo} 
configurations are favored at higher dark matter fractions, while the 
\textit{Core-Halo} configurations are more prevalent at lower 
$f_\mathrm{DM}$, as shown in the top-left panel of Fig.~\ref{halo_all}. In each top panel, the lighter-shaded bars represent the total number of M-R sequences belonging to that structural category, while the darker bars represent the observationally viable subset, i.e., those configurations that survive the applied constraints. Upon applying the combined GW+2.01+NICER $1\sigma$ constraint~\ref{cons4}, a clear difference in suppression is observed between the configurations. The \textit{Core-Halo} configurations (green) are effectively excluded beyond $f_{\mathrm{DM}} \gtrsim 0.10$. Similarly, the \textit{Halo} configurations (blue) survive only at lower dark matter fractions and are largely excluded at higher $f_{\mathrm{DM}}$, indicating that the combined tidal deformability and mass-radius constraints are highly effective in 
excluding high-$f_\mathrm{DM}$ Core-Halo and \textit{Halo} structures.

The bottom left panel presents the normalized distribution of the surviving M--R sequences for each configuration type as a function of $f_{\mathrm{DM}}$. The Core-Halo configurations are shown by the green dashed line with square markers, while the \textit{Halo} configurations are represented by the blue solid line with circular markers. These distributions quantify how the observational constraints restrict the allowed dark matter fraction for each configuration, demonstrating that both types are increasingly suppressed at higher $f_{\mathrm{DM}}$.

The middle panel of Fig.~\ref{halo_all} shows the distribution of M--R sequences for the dark matter model with $m_D = 0.7~\mathrm{GeV}$ across different configuration types. In contrast to the $m_D = 0.3~\mathrm{GeV}$ case, the \textit{Core} configuration (orange) dominates the parameter space, while \textit{Core-Halo} and \textit{Halo-Core} configurations appear primarily at intermediate and higher values of $f_{\mathrm{DM}}$.
Upon applying the combined GW+2.01+NICER $1\sigma$ constraint, a significant reduction in the number of M--R sequences is observed, particularly at higher $f_{\mathrm{DM}}$, where the \textit{Core} configurations are strongly suppressed. The \textit{Core-Halo} and \textit{Halo-Core} configurations are also considerably reduced, with only a small fraction of sequences remaining consistent with the observational constraints.
The bottom panel presents the corresponding normalized distribution as a function of $f_{\mathrm{DM}}$, clearly illustrating how the observational constraints progressively restrict the allowed dark matter fraction for each configuration type.

The rightmost panel of Fig.~\ref{halo_all} shows the distribution of M--R sequences for the dark matter model with $m_D = 1.0~\mathrm{GeV}$. In this case, the \textit{Core} configuration (orange) dominates across the entire range of $f_{\mathrm{DM}}$, while a small number of Halo-Core configurations appear at higher dark matter fractions. Upon applying the combined constraint~\ref{cons4}, the number of surviving \textit{Core} configurations is significantly reduced at higher $f_{\mathrm{DM}}$, indicating strong suppression by the observational constraints in this regime.

Overall, the statistical analysis across different dark matter models reveals a clear evolution in the dominant structural configurations as dark matter particle mass increases. For $m_D = 0.3~\mathrm{GeV}$, the M--R sequences are primarily composed of \textit{Core-Halo} and halo-dominated configurations, with halo structures becoming more prominent at higher $f_{\mathrm{DM}}$. In contrast, for $m_D = 0.7~\mathrm{GeV}$, the \textit{Core} configuration begins to dominate, while \textit{Core-Halo} and \textit{Halo-Core} configurations appear only at intermediate and higher dark matter fractions. At even larger mass, $m_D = 1.0~\mathrm{GeV}$, the parameter space is almost entirely dominated by \textit{Core} configurations, indicating a transition toward fully core-dominated structures.

The impact of observational constraints also varies systematically with $m_D$. The effects of the four different observational constraints (~\ref{astro_const}) on the DMANS configurations are presented in the Appendix \ref{app:config}. We show the normalized survival fraction of the M-R sequences as a function of the dark matter fraction, $f_{\mathrm{DM}}$, for each configuration type. For lower dark matter masses, halo-dominated configurations are strongly suppressed by tidal deformability constraints, whereas for higher $m_D$, where the dark matter distribution becomes increasingly core-like, the NICER mass-radius constraints play a more significant role in restricting the allowed parameter space. This trend highlights a clear shift in the dominant constraining mechanism, from tidal deformability measurements to mass-radius observations, as the dark matter distribution transitions from halo-dominated to core-dominated configurations.

\subsection{Distribution of DM-admixed neutron star configurations}

\begin{figure*}[ht]
    \centering
    \includegraphics[width=0.999\textwidth]{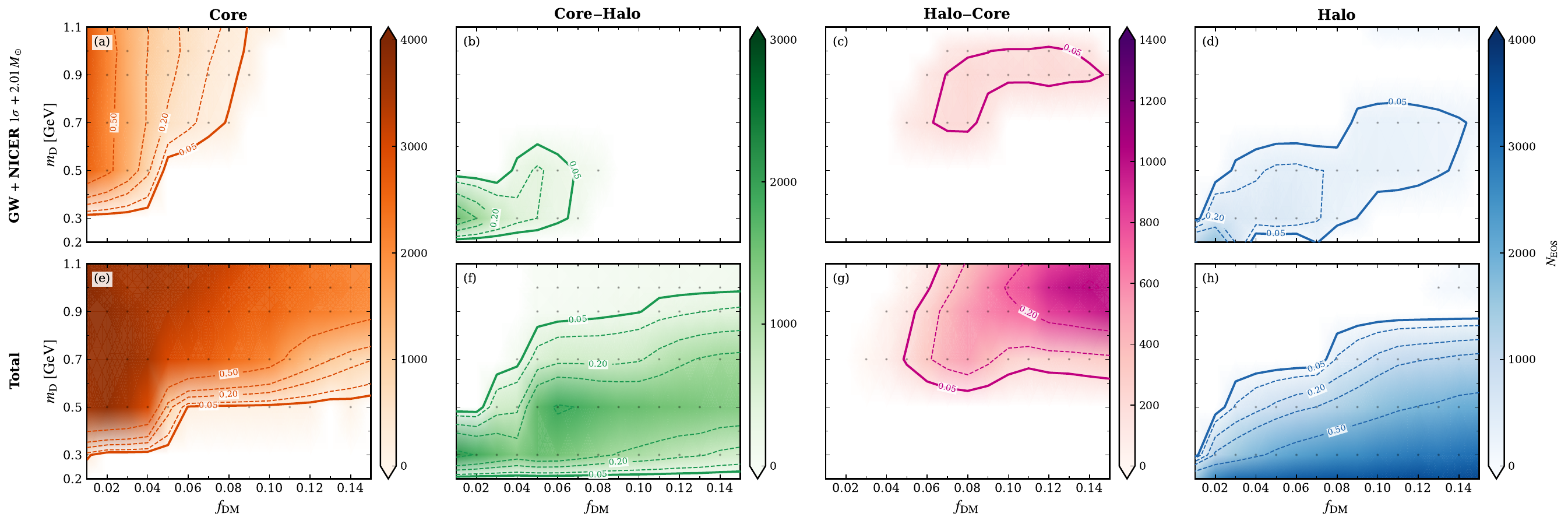}
    \caption{Distribution of dark matter-admixed neutron star configurations in the $(f_{\rm DM},\, m_{\rm D})$ parameter space. Columns correspond to the four configurations identified in this work: \textit{Core} (\textbf{left}), \textit{Core-Halo} (\textbf{middle-left}), \textit{Halo-Core} (\textbf{middle-right}) and \textit{Halo} (\textbf{right}). The top row shows EoS that satisfy current observational constraints (GW170817 tidal deformability, NICER $1\sigma$ mass-radius, and the $2.01\,M_\odot$ maximum-mass bound), while the bottom row shows the total EoS ensemble without observational filtering. Background color indicates the number of EoS, $N_{\rm EoS}$, falling into each $(f_{\rm DM},\, m_{\rm D})$ bin (see colorbar). Dashed contours mark iso-fractions $N_{\rm cat}/N_{\rm all}$ at levels $\{0.05,\,0.1,\,0.2,\,0.3,\,0.5\}$, indicating the fraction of EoS at that parameter point that produce the given configuration; the thicker solid contour denotes the $0.05$ level, taken as the boundary below which the configuration is effectively absent. Black dots mark grid cells with at least $N_{\rm EoS}^{\min}=10$ sampled EoS, where the statistics are considered reliable.}
\label{para_space}
\end{figure*}

\begin{figure*}
    \centering
    \begin{minipage}{0.49\linewidth}
        \centering
        \includegraphics[width=\linewidth]{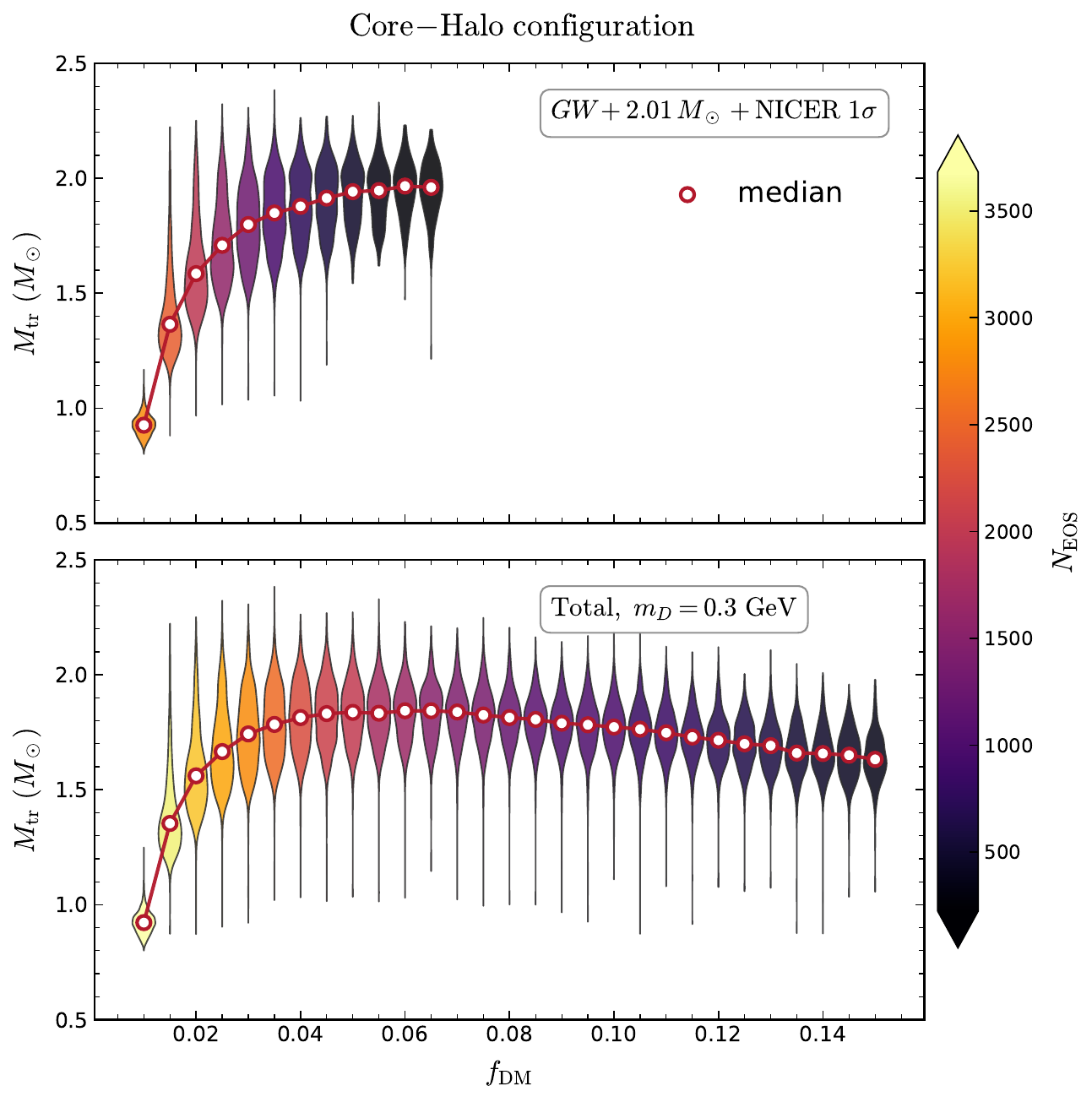}
    \end{minipage}
    \hfill
    \begin{minipage}{0.49\linewidth}
        \centering
        \includegraphics[width=\linewidth]{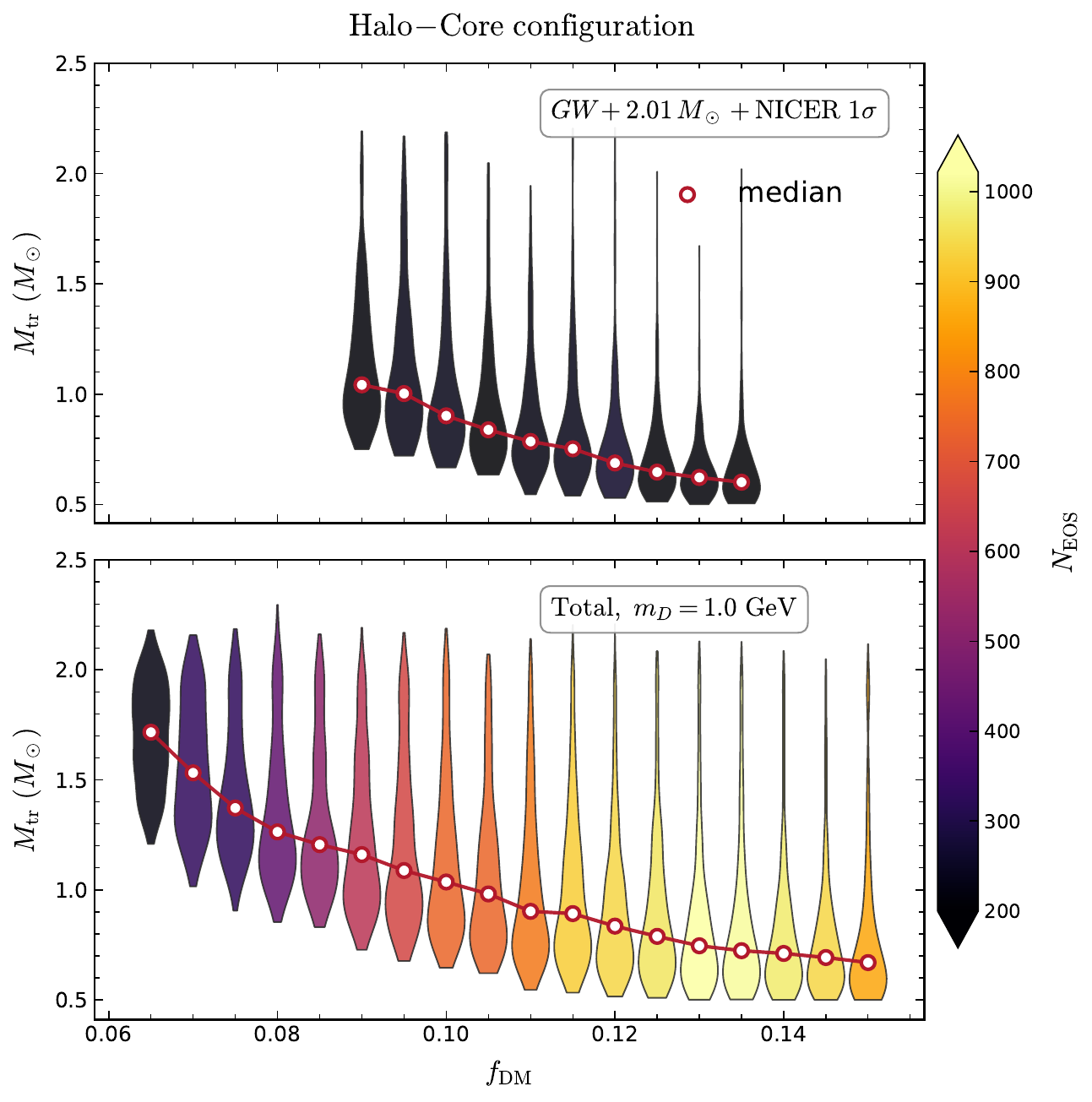}
    \end{minipage}
    \caption{Distribution of the Core-Halo transition mass, $M_{\mathrm{tr}}$, as a function of the dark matter fraction, $f_{\mathrm{DM}}$, for dark matter particle masses $m_D = 0.3$~GeV (\textbf{left panels}) and $m_D = 1.0$~GeV (\textbf{right panels}). The top panels show the distributions obtained from mass-radius sequences that satisfy both the gravitational-wave tidal deformability constraints and the NICER mass constraints, while the bottom panels show the distributions for the full set of sequences considered in each model. At each value of $f_{\mathrm{DM}}$, the violin plot represents the distribution of $M_{\mathrm{tr}}$ across all sampled equation-of-state combinations, with the width proportional to the kernel density estimate. The red curve traces the median transition mass. The color scale indicates the number of contributing equations of state, $N_{\mathrm{EoS}}$, at each dark matter fraction.}
    \label{m-trans}
\end{figure*}

Fig.~\ref{para_space} illustrates the distribution of the four DMANS 
configurations -- \textit{Core},  \textit{Core-Halo},  \textit{Halo-Core} and  \textit{Halo} -- in the parameter space ($f_{DM}, m_D$) defined by the dark matter fraction, $f_{\rm DM}$, and 
the dark matter particle mass, $m_{\rm D}$. The color scale represents the 
number of EoSs $N_{\rm EoS}$ that produce each 
configuration. The contour lines correspond to the iso-fraction
$Z=\frac{N_{\rm cat}}{N_{\rm all}}$,
where $N_{\rm cat}$ denotes the number of EoSs belonging to a given category 
and $N_{\rm all}$ is the total number of EoSs considered. The thick contour 
at $Z=0.05$ marks the boundary below which a configuration becomes 
statistically insignificant.

The lower panels (e--h) show the results obtained from the complete EoS 
ensemble, whereas the upper panels (a--d) display only those EoSs that 
satisfy the observational constraints from the GW170817 tidal deformability 
measurements, the NICER $1\sigma$ mass-radius observations, and the 
$2.01\,M_\odot$ maximum-mass requirement \ref{cons4}.

For the unconstrained EoS sample, a clear separation of the parameter space ($f_{DM}, m_D$) 
is observed. The DM \textit{Core} configuration, shown in panel~(e), is predominantly 
found at relatively large dark matter particle masses 
($m_{\rm D}\gtrsim0.5~{\rm GeV}$). In this regime, the heavy dark matter particles accumulate near 
the stellar center and form a compact dark matter core. In contrast, the DM 
\textit{Halo} configuration, shown in panel~(h), occupies the low-mass region 
($m_{\rm D}\lesssim0.7~{\rm GeV}$), especially at moderate and high 
DM fractions, where the dark matter extends beyond the visible 
stellar surface and forms a halo. The intermediate DM \textit{Core-Halo} and DM 
\textit{Halo-Core} configurations populate narrower transition regions between these 
two limiting cases. The DM Core-Halo configuration is primarily found for 
intermediate dark matter masses, spanning approximately 
$0.3 \lesssim m_{\rm D} \lesssim 0.9~{\rm GeV}$, whereas the DM Halo-Core 
configuration is concentrated at relatively large dark matter masses and high 
dark matter fractions.

The inclusion of observational constraints significantly modifies these distributions. For the DM \textit{Core} category, the allowed parameter space is 
substantially reduced and shifts towards lower values of $f_{\rm DM}$. 
High-fraction DM \textit{Core} configurations are strongly disfavored because a dense dark matter core softens the stellar structure and reduces the maximum 
mass below the observational limit. The DM \textit{Core-Halo} configuration is 
affected even more severely, with only a small region at low $m_{\rm D}$ and 
low $f_{\rm DM}$ remaining compatible with current observations, indicating 
that these transitional structures are particularly sensitive to the applied 
constraints.

The DM \textit{Halo-Core} configuration also survives only within a limited region 
at relatively large dark matter masses and high dark matter fractions. The DM 
\textit{Halo} configuration remains viable primarily at low dark matter masses and low 
dark matter fractions, while at larger dark matter masses, the surviving region 
shifts towards higher values of $f_{\rm DM}$. Although halo-dominated 
configurations are not excluded entirely, the number of compatible EoSs 
decreases significantly after applying the observational constraints. This 
behavior can be attributed to the fact that extended dark matter halos 
generally increase the stellar tidal deformability. Consequently, 
configurations with large dark matter fractions are strongly constrained by 
the GW170817 tidal-deformability bounds, leaving only low-fraction halo configurations observationally viable.

From Fig.~\ref{para_space}, it is evident that the DM model with $m_D = 0.2,\mathrm{GeV}$ does not exhibit \textit{Core}, \textit{Halo-Core}, or \textit{Core-Halo} configurations. Instead, all stable solutions correspond exclusively to the \textit{Halo} configuration, indicating that this model produces only halo-dominated DM distributions. In contrast, the DM model with $m_D = 1.1,\mathrm{GeV}$ yields only the \textit{Core} configuration, corresponding to purely core-dominated DM distributions.

These two limiting cases illustrate the transition in the internal DM structure as the DM particle mass increases. Models with $m_D \lesssim 0.2,\mathrm{GeV}$ produce only extended DM halos with very large DM radii. Such configurations are strongly disfavored by the tidal deformability constraints from gravitational-wave observations and are almost entirely excluded when combined with astrophysical constraints \ref{cons4}. On the other hand, models with $m_D \gtrsim 1.1,\mathrm{GeV}$ produce only compact DM cores. These configurations are primarily constrained by the NICER mass-radius observations, which restrict the allowed DM fraction because the compact DM core significantly alters the stellar mass-radius relation.

Motivated by these two extreme cases, we restrict our analysis to the intermediate mass range, $0.2 \lesssim m_D \lesssim 1.1,\mathrm{GeV}$, where all four DM configurations (\textit{Halo}, \textit{Halo-Core}, \textit{Core-Halo}, and \textit{Core}) can occur, allowing for a comprehensive investigation of the transition between halo-dominated and core-dominated neutron star configurations.

Overall, the figure highlights two important conclusions. First, the 
dark matter particle mass $m_{\rm D}$ is the primary parameter governing the 
internal dark matter structure of neutron stars: light dark matter favors 
halo formation, heavy dark matter favors core formation, and mixed 
configurations emerge in the intermediate regime. Second, current 
multimessenger constraints from GW170817, NICER, and the $2.01\,M_\odot$ maximum-mass measurement affect the four DMANS configurations differently. 
DM \textit{Core} configurations are predominantly excluded at high dark matter 
fractions, mainly due to the NICER mass-radius and maximum-mass constraints. 
The DM \textit{Core-Halo} and DM \textit{Halo-Core} configurations survive only in relatively 
small regions of parameter space. In contrast, DM \textit{Halo} configurations are 
primarily constrained by the GW170817 tidal-deformability bounds and the 
maximum-mass requirement, particularly at moderate and high dark matter 
fractions. These results demonstrate that present observational data provide 
strong discriminatory power between different dark matter distributions inside 
neutron stars and can significantly restrict the allowed DMANS parameter 
space.

It remains to be understood what the star mass is at the halo-core and core-halo transitions. To analyze this point, we consider two DM masses that predict each of these scenarios and plot the distribution of the transition mass as a function of the DM fraction.
From Fig. \ref{para_space}, we see that $m_{DM}=$0.3 GeV is a perfect example of \textit{Core-Halo}, and 1.0~GeV is a good example of \textit{Halo-Core} configurations. 
We considered these two values of the DM particle mass and plotted in Fig. \ref{m-trans} the distribution of the Core-Halo transition mass $M_{tr}$ for $m_{DM}=$0.3 GeV (left) and Halo-Core for $m_{DM}=$1.0 GeV (right), as a function of the dark matter fraction. The top (bottom) plots show the distributions when (no)  observational constraints are imposed.

For $m_{DM}=$0.3 GeV, the unconstrained data set allows for \textit{Core-Halo} configurations across all DM  fractions, with a distribution concentrated in masses above (below) 1.5$M_\odot$ for fractions above (below)  2\%. Imposing observational constraints limits the allowed region to DM fractions below 7\%. A larger number of configurations is found for $f_{DM}<0.03$ and NS masses between 1 and 1.5 $M_\odot$; however,  the halo effect can extend to quite high masses. The detection of two NS with similar masses but very different tidal deformabilities could be an indication of the presence of DM in one of the stars.

The distribution of \textit{Halo-Core} configurations obtained with $m_{DM}= 1.0$~GeV is shown in the right panels. These configurations occur only for large DM fractions ($>6\%$). Imposing observational constraints further reduces the allowed DM fraction interval to $9\%\lesssim f_{DM} \lesssim 13\%$. The median transition mass is always low ($\sim 0.5 - 1.0\, M_\odot$), although it can extend to large values.  The identification of this kind of configuration would be possible by detecting NS with similar tidal deformabilities but different masses, i.e., a high-mass star with a tidal deformability similar to that of a lower-mass star.

\section{Summary and Conclusions}
In this work, we have presented a comprehensive study of dark matter-admixed neutron stars (DMANSs) in which \emph{both} the nuclear-matter and the dark matter sectors are described by a single, agnostic speed-of-sound interpolation scheme. Whereas the agnostic reconstruction of the dense-matter EoS---bracketed by chiral effective field theory at low density and perturbative QCD at high density---is by now standard for nuclear matter, the dark sector has almost always been treated through a specific microscopic model. By extending the same speed-of-sound philosophy to the dark EoS (a free Fermi gas of bare mass $m_D$ below $0.1\,n_0$, followed by an agnostic, randomized $c_s^2$ profile with no high-density anchor), we are able to isolate the generic imprint of a dark component on neutron-star observables. In the future, we will analyze the dependence of the results on two assumptions considered for the DM EoS: i) the anchor density at low densities; ii) the fermionic description of the low-density DM EoS. The anchor density considered (0.1$n_0$)  falls in the NS crust. Therefore, we do not expect the main conclusions of the present work to be affected by these two assumptions and to be valid for the two-fluid hypothesis in general.
In future work, we also plan to investigate the low-density dependence of bosonic dark matter instead of the fermionic dark matter model considered in the present study.   %without committing to a bosonic, fermionic, or portal-coupled hypothesis.

Solving the two-fluid TOV and tidal equations for $\sim\!10^5$ M--R sequences per dark matter model, spanning $m_D = 0.2$--$1.1~\mathrm{GeV}$ and $f_{\mathrm{DM}} = 0.01$--$0.15$, and confronting them with the NICER mass-radius data, the GW170817 tidal bound $70 \leq \Lambda_{1.4} \leq 580$, and the $M_{\mathrm{max}} \geq 2.01\,M_\odot$ requirement, we arrive at the following main conclusions:
\begin{itemize}
\item dark matter generically compactifies the star at low and intermediate masses, shifting the M--R sequences toward smaller radii; for certain dark EoSs, however, the additional pressure support can instead stiffen the configuration and raise the maximum mass.
\item The dark matter distribution separates into core- and halo-dominated regimes that leave opposite imprints on the tidal deformability: extended halos enlarge the effective radius and increase $\Lambda$ (since $\Lambda \propto R^5$), whereas compact cores lower it.  Heavy mass DM particles give rise to both core (low DM fractions) and halo-core (high DM fractions). Measuring similar tidal deformability values for a low-mass and a high-mass star could identify halo-core configurations.
\item The particle mass $m_D$ controls this distribution: light dark matter ($m_D \lesssim 0.3~\mathrm{GeV}$) preferentially forms halos, heavy dark matter ($m_D \gtrsim 0.7~\mathrm{GeV}$) forms cores, and Core-Halo and Halo-Core sequences populate the intermediate regime.
\item The dominant observational constraint shifts accordingly: for light, halo-dominated models, the GW170817 tidal deformability is the most restrictive probe, limiting the permissible dark matter fraction to $f_{\mathrm{DM}} \lesssim 0.11$, whereas for heavy, core-dominated models, the NICER mass-radius measurements become the stronger constraint.
% \item Classifying the sequences by the sign of $dM/dR$, we find that current multi-messenger data systematically favor backward-bending over forward-bending configurations, the observational filtering removing the forward-bending-dominant region of the $(f_{\mathrm{DM}},\,m_D)$ plane almost entirely.
\end{itemize}

The observational constraints were taken with one $\sigma$ uncertainty in order to test whether it is possible to draw some conclusions. When considered at 2$\sigma$, the NICER constraining power was small. Taking a 1$\sigma$ uncertainty reflects the expected data from future observatories.

Taken together, these results show that present mass, radius, and tidal-deformability measurements may already place meaningful bounds on the dark matter content of neutron stars, and that the relative constraining power of the X-ray and gravitational-wave observations is itself a diagnostic of the underlying dark matter distribution. Because our framework makes almost no assumptions about the microphysics of the dark sector, the bounds obtained here are conservative and broadly applicable. Future NICER mass-radius measurements of additional pulsars, together with tidal-deformability constraints from next-generation gravitational-wave detectors, are expected to sharpen these limits considerably and may ultimately help distinguish halo- from core-dominated dark matter-admixed neutron stars.

\section{acknowledgements}
The authors, A.K. and R.M., thank the Indian Institute of Science Education and Research Bhopal for providing all the research and infrastructure facilities.
R.M. acknowledges support from the Science and Engineering Research Board (SERB), Govt. of India, for financial support in the form of a Core Research Grant (CRG/2022/000663). This work was supported by computational resources from the Bhaskara and Gargi HPC systems in IISER Bhopal. T.M. and R.M. express gratitude for the Deucalion HPC platform in Portugal, appreciating its resources and technical support, facilitated by the FCT project 2025.00067.CPCA.A3. T.M. and C.P.
received support from Fundação para a Ciência e a Tecnologia
(FCT), I.P., Portugal, under the project
 UID/04564/2025, identified by DOI 10.54499/UIDB/04564/2025 and the project 2024.16290.PEX with DOI identifier 10.54499/2024.16290.PEX.k \\

\section{DATA AVAILABILITY}
The data that support the findings of this article are not publicly available upon publication because it is not technically feasible, and/or the cost of preparing, depositing, and hosting the data would be prohibitive within the terms of this research project. The data are available from the authors upon reasonable request.

% \bibliographystyle{apsrev}
% % \bibliographystyle{abbrv}
% \bibliography{refs}

\appendix

\section{Configuration-resolved survival fractions}
\label{app:config}
\begin{figure*}
    \centering
    \includegraphics[width=1.0\textwidth]{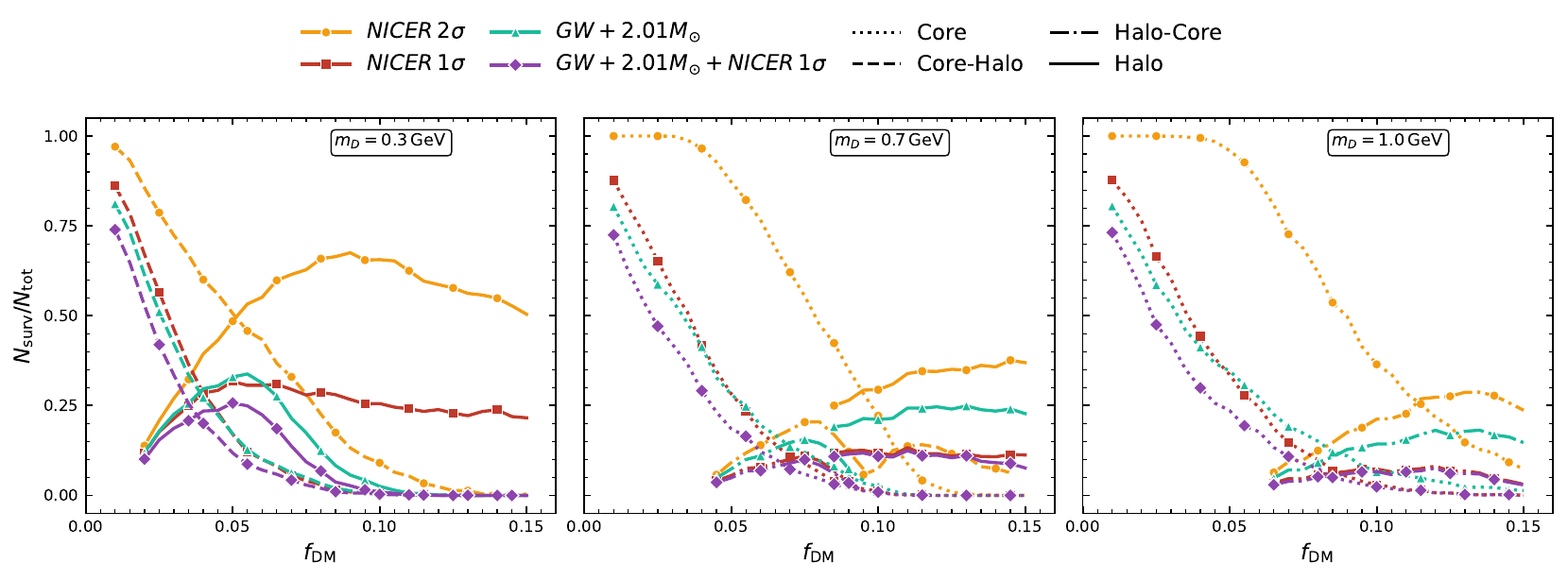}
    \caption{normalized survival fraction, $N_{\mathrm{surv}}/N_{\mathrm{tot}}$, of M--R sequences for each configuration type as a function of dark matter fraction, $f_{\mathrm{DM}}$, under four observational constraint sets: NICER $2\sigma$ (orange, circles), NICER $1\sigma$ (red, squares), GW+2.01 (cyan, triangles), and GW+2.01+NICER $1\sigma$ (purple, diamonds). The four configuration types are distinguished by line style: Halo (solid), Core-Halo (dashed), Core (dotted), and Halo-Core (dash-dotted). Results are shown for (\textbf{Left}) $m_D = 0.3\,\mathrm{GeV}$, (\textbf{Middle}) $m_D = 0.7\,\mathrm{GeV}$, and (\textbf{Right}) $m_D = 1.0\,\mathrm{GeV}$.}
    \label{halo_stat}
\end{figure*}

Fig.~\ref{halo_stat} complements Sec .~\ref {core_halo_config} by resolving the normalized survival fraction according to the structural configuration type of each M--R sequence.

To examine the relative impact of the tidal-deformability and mass-radius constraints on the different DMANS configurations, Fig.~\ref{halo_stat} shows the normalized survival fraction of M--R sequences as a function of $f_{\mathrm{DM}}$ for each configuration type, under the four observational constraint sets defined in Sec.~\ref{astro_const} (NICER $2\sigma$, NICER $1\sigma$, GW+2.01, and the combined GW+2.01+NICER $1\sigma$).

The left panel corresponds to $m_D = 0.3~\mathrm{GeV}$. For halo-dominated configurations (\textit{Halo}, solid line) the tidal-deformability constraint (GW+2.01) is more effective at suppressing the extended-halo configurations than the NICER mass-radius constraints. In contrast, for \textit{Core-Halo} (dashed line) configurations, the NICER $1\sigma$ and GW+2.01 constraints are comparably effective.

The middle panel shows $m_D = 0.7~\mathrm{GeV}$. Here, the \textit{Core} configurations (dotted line) dominate, and both the tidal and the mass-radius constraints limit the parameter space with similar strength; for \textit{Halo-Core} configurations (dash-dotted line), the NICER $1\sigma$ and GW+2.01 constraints again provide comparable suppression across the full $f_{\mathrm{DM}}$ range.

The right panel presents $m_D = 1.0~\mathrm{GeV}$, where the parameter space is dominated by \textit{Core} configurations with only a small fraction of Halo-Core structures. For the \textit{Core} configurations, the two constraints remain broadly comparable, although at higher $f_{\mathrm{DM}}$, the NICER mass-radius constraint becomes the more effective one; a similar trend holds for the Halo-Core configurations.

Overall, the relative effectiveness of the tidal deformability and mass-radius constraints depends sensitively on the underlying dark matter distribution: halo-dominated structures are more strongly constrained by the tidal deformability, whereas core-dominated configurations are constrained with comparable strength by both observations, with a slight preference for the mass-radius constraints at higher dark matter fractions.

\begin{figure*}[!t]
    \centering
    \includegraphics[width=1.0\textwidth]{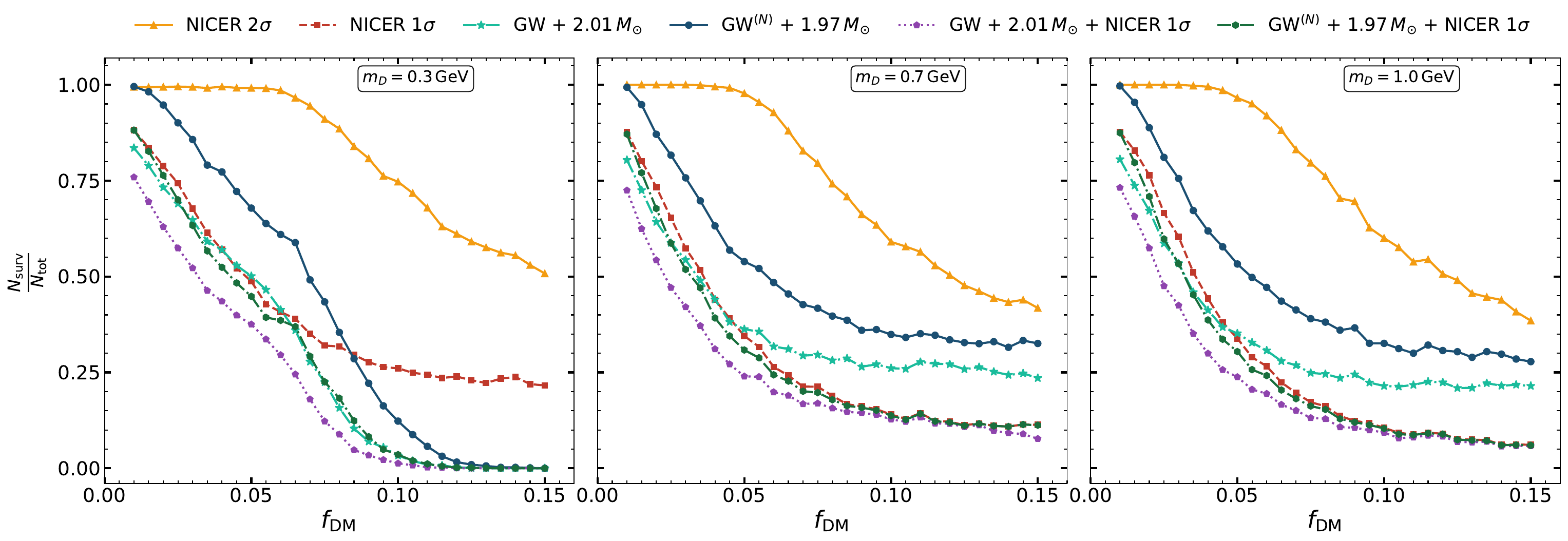}
    \caption{ Same as Fig. \ref{norm_all}, but including the broader GW170817 tidal deformability constraint, $\mathrm{GW}^{(N)}$ ($70 \leq \Lambda \leq 800$), in addition to the maximum mass constraint $M_{\rm max} \geq 1.97,M_\odot$, and the combined constraint with the NICER $1\sigma$ measurements. The normalized survival fraction of M--R sequences, $N_{\mathrm{surv}}/N_{\mathrm{tot}}$, is shown as a function of the dark matter fraction, $f_{\mathrm{DM}}$, for (\textbf{Left}) $m_D = 0.3,\mathrm{GeV}$, (\textbf{Middle}) $m_D = 0.7,\mathrm{GeV}$, and (\textbf{Right}) $m_D = 1.0,\mathrm{GeV}$.}
    \label{norm_atro_upd}
\end{figure*}

\section{Distribution of Permissible Dark Matter Fractions under Broader Astrophysical Constraints}
\label{broad_apc}
Fig.~\ref{norm_atro_upd} extends the analysis presented in Sec.~\ref{fdm_const} by examining how broader astrophysical constraints affect the distribution of permissible DM fractions in DMANSs. In addition to the reference constraints used previously, we consider two alternative combinations:
(i) $GW^{(N)} + 1.97$, where the tidal deformability constraint from GW170817 is relaxed to $70 \leq \Lambda_{1.4} \leq 800$, together with the maximum mass requirement $M_{\rm max}\geq1.97\,M_\odot$, and
(ii) $GW^{(N)} + 1.97 + \mathrm{NICER}\;1\sigma$, where the above constraints are further combined with the NICER $1\sigma$ mass-radius measurements.

The left panel of Fig.~\ref{norm_atro_upd} presents the normalized survival fraction of M--R sequences as a function of the DM fraction for the $m_D=0.3$ GeV model. Compared with the reference constraint ($GW + 2.01$, cyan star), the broader constraint ($GW^{(N)} + 1.97$, blue circles) is slightly less restrictive, allowing a larger fraction of configurations to survive over the same range of DM fractions. Nevertheless, both constraints exhibit nearly identical trends, with the survival fraction decreasing rapidly as $f_{\rm DM}$ increases and vanishing at sufficiently large DM fractions. This demonstrates that the tidal deformability constraint, even in its relaxed form, remains highly effective in excluding large DM fractions for low-mass DM models. The addition of the NICER $1\sigma$ constraint (dark green hexagons) further suppresses the surviving configurations, producing a distribution very similar to that obtained using the $GW + 2.01 + \mathrm{NICER}\;1\sigma$ constraint and restricting the allowed DM fraction to approximately $f_{\rm DM}\lesssim0.11$.

The middle and right panels show the corresponding distributions for the $m_D=0.7$ GeV and $m_D=1.0$ GeV models, respectively. The broader constraint $GW^{(N)} + 1.97$ again yields a slightly higher survival fraction than the reference $GW + 2.01$ constraint while preserving the same overall dependence on the DM fraction. Including the NICER $1\sigma$ constraint further reduces the number of surviving configurations and shifts the distributions toward lower DM fractions, closely matching the behavior obtained with the corresponding reference constraint.

It should be noted that the combined constraint $GW^{(N)} + 1.97 + \mathrm{NICER}\;2\sigma$ is not considered in the present analysis. Since the NICER $2\sigma$ constraint is comparatively weak in constraining the DM fraction, its combination with the more restrictive $GW^{(N)} + 1.97$ constraint yields essentially the same distribution of surviving M--R sequences as obtained from the $GW^{(N)} + 1.97$ constraint alone. Consequently, this combination provides no additional constraining power and is therefore omitted.

Overall, adopting the broader astrophysical constraints results in only modest quantitative changes to the survival fraction. While the relaxed tidal deformability and maximum mass constraints permit a slightly larger region of parameter space, they do not alter the qualitative conclusions. High DM fractions remain strongly disfavored for all three DM models, demonstrating that the inferred upper limits on the permissible DM fraction are robust against reasonable variations in the adopted astrophysical constraints.

%\color{blue}
\section{Dependence of DMANS Configuration on Anchor Density}
\label{anchor_pnt}
\begin{figure}[ht]
    \centering
    \includegraphics[width=0.45\textwidth]{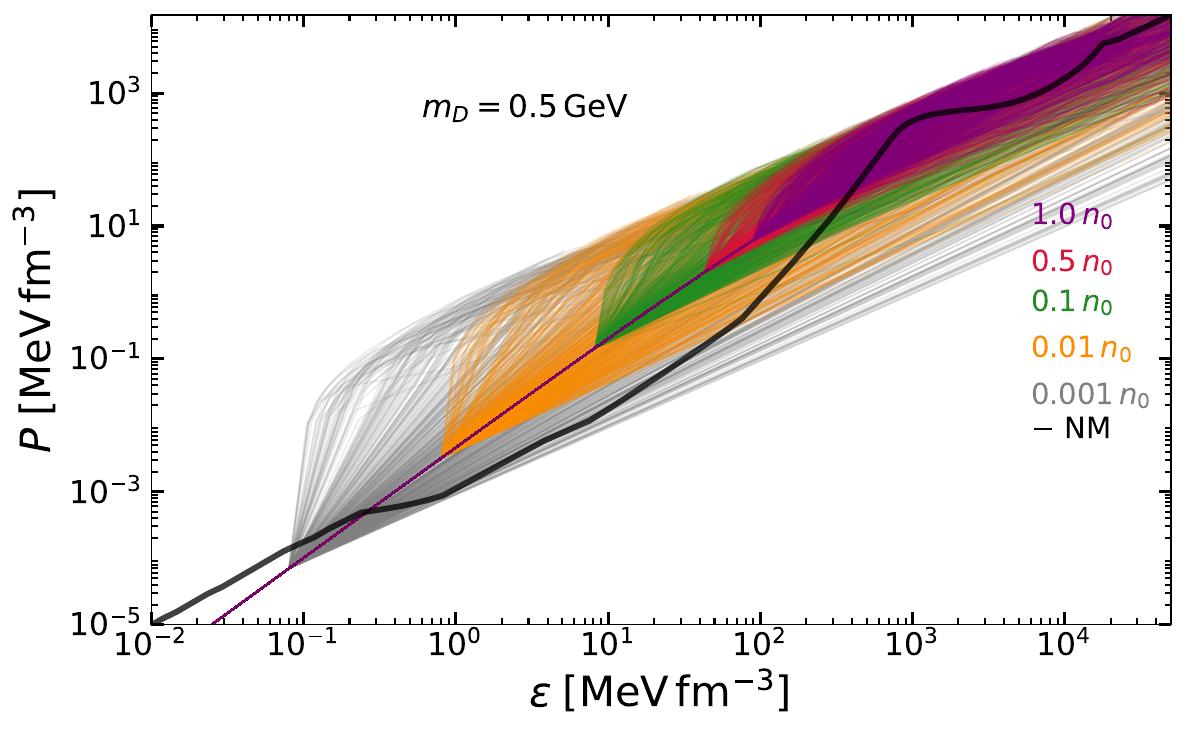}
    \caption{Agnostic DM EoSs constructed using different anchor densities, $0.001n_0$ (gray), $0.01n_0$ (orange), $0.1n_0$ (green), $0.5n_0$ (red), and $1.0n_0$ (purple), along with the NM EoS (solid black curve).}
    \label{fig:anchor_density}
\end{figure}

To investigate the dependence of our results on the choice of anchor density and assess its impact, we construct agnostic DM EoSs using five different anchor densities, namely, \(0.001n_0\), \(0.01n_0\), \(0.1n_0\), \(0.5n_0\), and \(1.0n_0\), for a fixed DM particle mass of \(m_D=0.5\) GeV, as shown in Fig.~\ref{fig:anchor_density}.
To isolate the effect of the anchor density on DMANS configurations, we fix the NM EoS to a representative EoS that is already well constrained by current astrophysical observations, shown by the solid black curve in the figure. We then construct DMANS M--R sequences for a fixed dark matter fraction, $f_{\rm DM}=0.15$, for each of the five anchor densities. For each anchor density, we construct 500 $M$--$R$ sequences, corresponding to 500 different realizations of the DM EoS.

\begin{figure*}[ht!]
    \centering

    \begin{minipage}{0.48\textwidth}
        \centering
        \includegraphics[width=\textwidth]{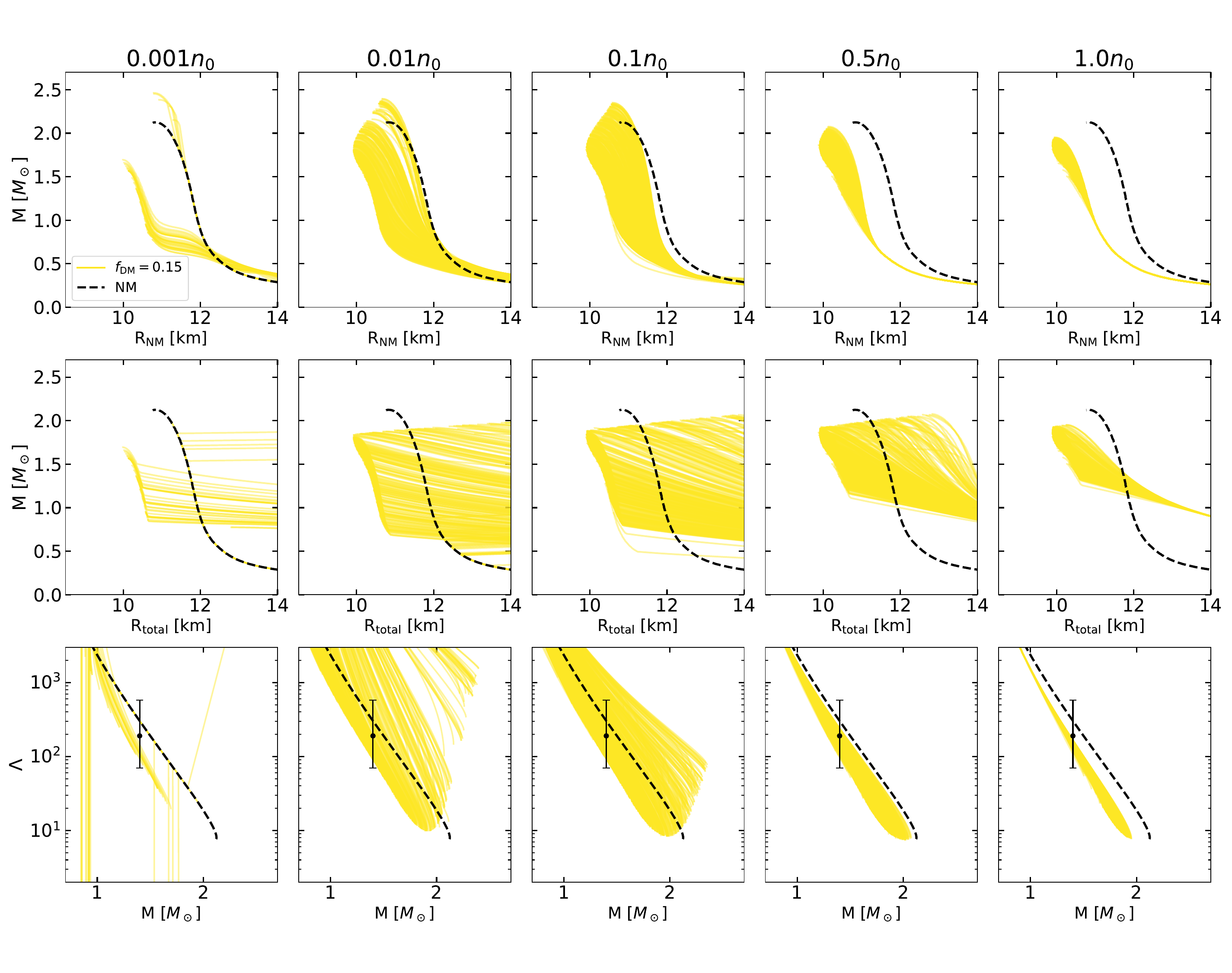}
    \end{minipage}
    \hfill
    \begin{minipage}{0.48\textwidth}
        \centering
        \includegraphics[width=\textwidth]{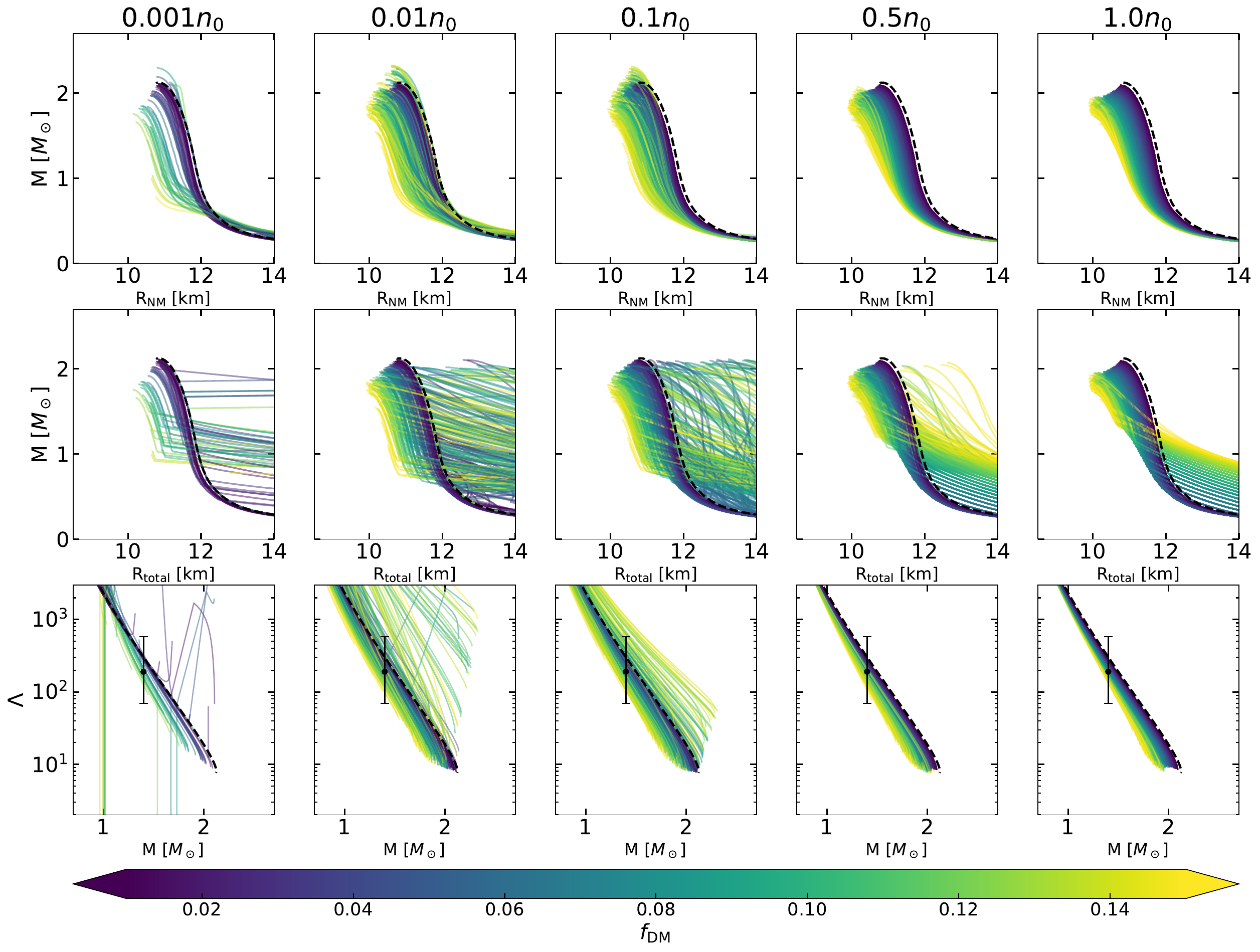}
    \end{minipage}

    \caption{mass-radius (M--R) and tidal deformability relations for different anchor densities, $n_{\rm anchor}=0.001n_0$, $0.01n_0$, $0.1n_0$, $0.5n_0$, and $1.0n_0$, considering a fixed dark matter fraction $f_{\rm DM}=0.15$ (left) and a varying dark matter fraction $0.01\leq f_{\rm DM}\leq0.15$ (right).}
    \label{MR}
\end{figure*}

The left panel of Fig.~\ref{MR} shows the resulting $M$--$R$ relations and tidal deformabilities for the five sets of DM models. The top row shows the $M$--$R$ relations using the nuclear matter radius, while the middle row shows the $M$--$R$ relations using the total radius ($R_{total}$), including the DM component. The bottom row presents the corresponding tidal deformability, $\Lambda$, for these configurations.

It is evident from the figure that, for the higher anchor densities, $n_{\rm anchor}=1.0n_0$ and $0.5n_0$, the $M$--$R$ relations become nearly identical at lower stellar masses. For the $1.0n_0$ case, this behavior occurs below approximately $1M_\odot$, whereas for the $0.5n_0$ case, it extends to approximately $0.75 M_\odot$. This behavior arises because, in these mass ranges, essentially the same DM EoS is sampled across the 500 realizations, as can also be understood from the corresponding DM EoS shown in Fig.~\ref{fig:anchor_density}

For $n_{\rm anchor}=1.0n_0$, only core-dominated configurations are obtained, as evident from the middle row of the figure. In contrast, for $n_{\rm anchor}=0.5n_0$, a number of configurations with smaller-radius DM halos are also obtained. Consequently, the tidal deformability is severely constrained in the higher-anchor-density cases, reflecting the smaller effective stellar radii associated with these configurations. Thus, the basic idea of the agnostic approach and obtaining a significant ensemble of thermodynamically consistent EoS fails.

When considering an anchor density close to $0.1n_0$, we find that the resulting models cover a broad range of physically viable configurations, including both core-dominated and halo configurations. In addition, the lower-mass stars exhibit a wide variation in their $M$-$R$ relations due to the different DM EoS realizations. Some of these realizations also produce configurations with masses larger than those of the corresponding DM-free neutron stars. The resulting distribution of tidal deformabilities is shown in the bottom panel of Fig.~\ref{MR}.

The primary goal of our analysis is to construct a general ensemble of DM EoSs that spans a broad range of possible DM-matter behaviors for a given DM particle mass. We therefore require an anchor density that provides a sufficiently broad range of physically viable high-density DM EoSs while allowing us to efficiently construct DM-admixed neutron-star sequences. Based on the above analysis, we find that $0.1n_0$ provides a suitable choice of anchor density for constructing the high-density DM EoS using the randomized speed-of-sound scheme.

We also consider lower anchor densities, namely, $0.01n_0$ and $0.001n_0$. For the $0.01n_0$ anchor, we find that the resulting models do produce a broad range of configurations, including both core and halo configurations, similar to those obtained for the $0.1n_0$ anchor. However, this choice also generates significant unphysical $M$--$R$ configurations due to the lower anchor density, which are subsequently rejected. Therefore, in order to construct a sufficiently large sample of physically viable dark matter-admixed neutron-star sequences, we need to generate a larger number of DM EoS realizations to obtain the same number of physical $M$--$R$ sequences as obtained for the $0.1n_0$ anchor. This makes the calculation numerically more expensive, without gaining any new qualitative results.

If we further decrease the anchor density to $0.001n_0$, the number of physically viable $M$--$R$ sequences becomes considerably smaller for the same number of DM EoS realizations. In addition, some of these EoSs exhibit rather unusual tidal responses. Thus, the efficiency of obtaining physically viable neutron-star configurations decreases further as the anchor density is lowered.

As evident from Fig.~\ref{fig:anchor_density}, although lower anchor densities can cover a broad range of the $(P,\epsilon)$ parameter space, obtaining the same number of physically viable $M$--$R$ sequences requires us to sample a substantially larger EoS parameter space than that for the higher anchor density. Therefore, one may in principle adopt lower anchor densities, such as $0.001n_0$ or $0.01n_0$, but doing so is computationally more expensive, while the generic features of the resulting DM-admixed neutron-star configurations remain qualitatively similar.

We have also examined this behavior for different dark-matter fractions, ranging from $f_{\rm DM}=0.01$ to $0.15$, using the same 500 DM EoS realizations for each of the five anchor densities. The corresponding results, shown in the right panel of Fig.~\ref{MR}, further support the above conclusion. We have verified that this analysis remains valid for other dark-matter particle masses, $m_D$.

\bibliographystyle{apsrev}
\bibliography{refs}

\end{document}